\documentclass[preprint,showpacs,preprintnumbers]{revtex4}
\usepackage[latin9]{inputenc}
\usepackage{amsmath}
\usepackage{amssymb}
\usepackage{graphicx}
\usepackage{esint}
\usepackage{times}
\usepackage{dcolumn}
\usepackage{bm}
\usepackage{latexsym}

\makeatletter
\@ifundefined{textcolor}{}
{
 \definecolor{BLACK}{gray}{0}
 \definecolor{WHITE}{gray}{1}
 \definecolor{RED}{rgb}{1,0,0}
 \definecolor{GREEN}{rgb}{0,1,0}
 \definecolor{BLUE}{rgb}{0,0,1}
 \definecolor{CYAN}{cmyk}{1,0,0,0}
 \definecolor{MAGENTA}{cmyk}{0,1,0,0}
 \definecolor{YELLOW}{cmyk}{0,0,1,0}
}
\@ifundefined{textcolor}{}{
 \definecolor{BLACK}{gray}{0}
 \definecolor{WHITE}{gray}{1}
 \definecolor{RED}{rgb}{1,0,0}
 \definecolor{GREEN}{rgb}{0,1,0}
 \definecolor{BLUE}{rgb}{0,0,1}
 \definecolor{CYAN}{cmyk}{1,0,0,0}
 \definecolor{MAGENTA}{cmyk}{0,1,0,0}
 \definecolor{YELLOW}{cmyk}{0,0,1,0}
}
\makeatother

\begin{document}

\title{Theory of inelastic multiphonon scattering and carrier capture by
defects in semiconductors -- Application to capture cross sections }
\thanks{This manuscript has been authored by UT-Battelle, LLC under Contract
No. DE-AC05-00OR22725 with the U.S. Department of Energy. The United States
Government retains and the publisher, by accepting the article for
publication, acknowledges that the United States Government retains a
non-exclusive, paid-up, irrevocable, world-wide license to publish or
reproduce the published form of this manuscript, or allow others to do so,
for United States Government purposes. The Department of Energy will provide
public access to these results of federally sponsored research in accordance
with the DOE Public Access
Plan(http://energy.gov/downloads/doe-public-access-plan).}
\author{Georgios D. Barmparis$^{1}$, Yevgeniy S. Puzyrev$^{1}$, X.-G. Zhang$%
^{2}$ and Sokrates T. Pantelides$^{1,3,4}$}
\affiliation{$^{1}$Department of Physics and Astronomy, Vanderbilt University, Nashville,
Tennessee, 37235\\
$^{2}$Department of Physics and the Quantum Theory Project, University of
Florida, Gainesville, Florida 32611\\
$^{3}$Materials Science and Technology Division, Oak Ridge National
Laboratory, Oak Ridge, Tennessee, 37831\\
$^{4}$Department of Electrical Engineering and Computeer Science, Vanderbilt
University, Nashville, TN\ 37235}

\begin{abstract}
Inelastic scattering and carrier capture by defects in semiconductors are
the primary causes of hot-electron-mediated degradation of power devices,
which holds up their commercial development. At the same time, carrier
capture is a major issue in the performance of solar cells and
light-emitting diodes. A theory of nonradiative (multiphonon) inelastic
scattering by defects, however, is non-existent, while the theory for
carrier capture by defects has had a long and arduous history. Here we
report the construction of a comprehensive theory of inelastic scattering by
defects, with carrier capture being a special case. We distinguish between
capture under thermal equilibrium conditions and capture under
non-equilibrium conditions, e.g., in the presence of electrical current or
hot carriers where carriers undergo scattering by defects and are described
by a mean free path. In the thermal-equilibrium case, capture is mediated by
a non-adiabatic perturbation Hamiltonian, originally identified by Huang and
Rhys and by Kubo, which is equal to linear electron-phonon coupling to first
order. In the non-equilibrium case, we demonstrate that the primary capture
mechanism is within the Born-Oppenheimer approximation (adiabatic
transitions), with coupling to the defect potential inducing Franck-Condon
electronic transitions, followed by multiphonon dissipation of the
transition energy, while the non-adiabatic terms are of secondary importance
(they scale with the inverse of the mass of typical atoms in the defect
complex). %Overall, the capture cross section
%depends on the elastic mean free path and the capture mean free path. 
We report first-principles density-functional-theory calculations of the
capture cross section for a prototype defect using the
Projector-Augmented-Wave which allows us to employ all-electron
wavefunctions. We adopt a Monte Carlo scheme to sample multiphonon
configurations and obtain converged results. The theory and the results
represent a foundation upon which to build engineering-level models for
hot-electron degradation of power devices and the performance of solar cells
and light-emitting diodes.
\end{abstract}

\date{\today }
\maketitle

\section{Introduction}

Elastic scattering of electrons by phonons, impurities, and other defects
limits the conductivity in metals and the carrier mobility in
semiconductors. The fundamental theory is well established, parameter-free
mobility calculations have become possible \cite{Evans,Hadjisavvas}, and
engineering-level modeling methods are widely available. Inelastic
scattering of hot electrons by defects has long been known to cause device
degradation. For example, hot electrons in Si-SiO$_{2}$ structures can
transfer energy and release hydrogen from passivated interfacial Si dangling
bonds \cite{DiMaria:1999,DiMaria:2000}. More recently, it was found that hot
electrons cause degradation of power devices based on wide-band-gap
semiconductors \cite{Meneghesso}. It has been shown that the degradation is
caused by hot-electron-mediated release of hydrogen from hydrogenated
defects such as Ga vacancies or impurities \cite{Pantelides:2012}. In other
cases, carrier capture transforms benign defects to metastable
configurations that cause recoverable degradation \cite{Shen}. Similarly,
non-radiative carrier capture by defects, which is a special case of
inelastic scattering, limits the performance of photovoltaic cells,
light-emitting diodes and other devices \cite{Meneghini,Igalson}.

A theory of inelastic scattering by defects by multiphonon processes (MPPs)
does not exist while the theory of non-radiative carrier capture or emission
by defects\ by MPPs has a long and controversial history. In 1950, Huang and
Rhys \cite{Huang:1950} reported a theory of how the energy of lattice
relaxation that accompanies the photoionization of a defect is dissipated by
MPPs. The process was described within the Born-Oppenheimer or adiabatic
approximation (BOA) and the Frank-Condon approximation (FCA). The former
says that the electronic and nuclear (vibrational) wave functions obey
decoupled equations. The latter states that an electronic excitation occurs
instantaneously and relaxation processes follow at a relatively slow pace,
allowing one to write the excitation rate (Fermi's golden rule) as a product 
$P=AF$, where $A$ describes the instantaneous electronic excitation in the
initial lattice configuration and $F,$ the so-called line-shape function,
describes the MPPs\ that occur during lattice relaxation. In the Huang-Rhys
theory, the operator that causes the excitation is strictly the photon field
and MPPs dissipate only the energy of the ensuing lattice relaxation.

In the same paper, Huang and Rhys \cite{Huang:1950} also proposed a theory
for non-radiative multiphonon transitions between defect levels. Such
transitions are caused by the terms that are dropped when the
Born-Oppenheimer approximation (BOA) is made, namely derivatives of the
electronic wavefunctions with respect to nuclear positions (non-adiabatic
terms). In 1952, Kubo \cite{Kubo:1952} independently invoked the same
non-adiabatic terms as being responsible for the thermal ionization of a
defect. In subsequent years, Kubo and Toyozawa \cite{Toyozawa:1955} and
later Gummel and Lax \cite{Gummel:1957} adopted Kubo's formalism to explore
carrier capture and emission using analytical approximations. Kovarskii and
Sinyavskii \cite{Kovarskii:1962,Kovarskii:1963,Kovarskii:1967} published
several papers expanding on Kubo's formalism. In 1977, in search of a
practical scheme to model electron capture in experiments , Henry and Lang 
\cite{Henry:1977} adopted a Huang-Rhys analog: the electronic transition is
caused instantaneously by the perturbation potential $\Delta V$ generated by
atomic vibrations -- the linear electron-phonon coupling potential that is
normally thought to cause elastic scattering and is used for mobility
calculations. The following year, Ridley \cite{Ridley:1978} showed that the
Henry-Lang model exhibits the correct temperature dependence at high
temperatures (the semi-classical limit), but pointed out that the correct
way to calculate non-radiative capture cross sections is through the
non-adiabatic perturbation terms identified by Huang and Rhys \cite%
{Huang:1950} and by Kubo \cite{Kubo:1952}. In 1981, however, Huang showed
that the non-adiabatic perturbation Hamiltonian and the linear
electron-phonon coupling perturbation Hamiltonian are equivalent to first
order \cite{Huang:1981}. The issue whether such a first-order calculation is
adequate remained open as, throughout the years of all these developments,
only model calculations were pursued, largely analytical, employing model
defect wave functions. Furthermore, calculations of the line-shape function
were typically restricted by the assumption that a single vibrational mode
contributes to the MPPs. In the chemical literature, noradiative transitions
between molecular orbitals have been studied \cite{Doktorov,Borrelli}. It
was recognized that inclusion of all vibrational modes in the MPP\
calculation leads to exploding computational requirements as the size of the
molecule increases \cite{Doktorov}. The so-called parallel-mode
approximation or simply a single vibrational mode are typically used \cite%
{Borrelli}.

The first application of modern density-functional-theory (DFT) calculations
to MPPs in the case of luminescence, i.e., the classic Huang-Rhys problem
where an electronic transition is caused by the photon field and MPPs
dissipate the ensuing lattice relaxation, was reported by Alkauskas et al. 
\cite{REF2012}. These authors studied the luminescence spectra of defects in
GaN employing DFT\ pseudo wave functions for the electronic matrix elements
and the single-phonon-mode aproximation to the Huang-Rhys line-shape
function. In a more recent paper, Alkauskas et al. \cite{Alkauskas} reported
calculations of non-radiative capture of carriers by defects using the
linear electron-phonon coupling perturbation Hamiltonian, pseudo wave
functions, and a single-phonon-mode to calculate the MPPs that dissipate the
transition energy. They pointed out that the electronic transition is a slow
process because capture is mediated by the phonons that are localized around
the defect.

In this paper we first revisit the theory of carrier capture by defects. We
identify two distinct regimes that are governed by different processes. One
is carrier capture under thermal equilibrium conditions, i.e., capture
occurs in tandem with emission and electrons in the conduction band (or
holes in the valence band) are not being accelerated. Under these
conditions, capture and emission are inverse processes, i.e., the role of
the initial and final states is reversed. For an electron bound at a defect,
emission amounts to a transition to a band state that is an eigenstate of
the same Hamiltoninan (perfect crystal plus defect potential). Band states
are occupied according to the Fermi-Dirac distribution function. Any of
these carriers can be captured into the defect's ground state. Under such
conditions, band carriers are effectively undergoing diffusive Brownian
motion. In this case, the Huang-Rhys-Kubo (HRK) non-adiabatic Hamiltonian
perturbation is the only possible cause for these thermal transitions.

Under non-equilibrium conditions, however, e.g., in the presence of an
electrical current, carriers are accelerated in a specific direction and a
mean free path is defined by scattering events. It is then standard
procedure to treat the band electrons as being in eigenstates of the perfect
crystal Hamiltonian and consider scattering by the defects. In particular,
one considers elastic scattering by defects as a mechanism that limits the
carrier mobility. In this case, the initial and final states are eigenstates
of the perfect crystal Hamiltonian and the defect potential acts as the
perturbation that causes the transitions, i.e., the defect potential is
``turned on'' in order to use time-dependent perturbation theory and arrive
at Fermi's golden rule. Clearly, hot carriers can undergo inelastic
scattering as well, dropping to a Bloch state of lower energy, with the
energy dissipated by MPP. For such calculations, one must again ``turn on''
the defect potential, though the HRK non-adiabatic perturbation must also be
included. Transitions caused by the defect potential are within the BO\
approximation, whereas those caused by the HRK\ perturbation Hamiltonian are
non-adaiabatic. Finally, under such non-equilibrium conditions, carrier
capture can be viewed as a special case of inelastic scattering: if the
defect potential can cause elastic scattering and inelastic scattering with
energy dissipation via MPP, then it certainly should also be included as a
cause for capture.

In the capture case, however, there is a subtle difficulty. In order to
derive a transition rate using Fermi's golden rule, initial and final states
must be eigenstates of the same Hamiltonian. In the carrier capture case,
however, the final state is an eigenstate of the crystal Hamiltonian plus
the defect potential, whereas the initial state is an eigenstate of the
perfect crystal Hamiltonian. The difficuty can be overcome if we prepare a
propagating state for the incoming electron that is not aware of the bound
state's existence, with capture being triggered by the sudden turning on of
a suitable coupling (initial and final states must belong to the same
Hamiltonian for the concept of a transition to be meaningful) to the defect
potential. Such adiabatic transitions have not been considered so far in the
context of multiphonon transitions at defects in semiconductors, but they
are commonly invoked in chemistry for elecron transitions in molecules \cite%
{MayBook,HUSH1958,HUSH1961}.

We will develop a comprehensive theory of inelastic scattering and capture
for transitions caused by both the defect potential (adiabatic transitions)
and by the non-adiabatic HRK perturbation Hamiltonian. We will show that,
for carrier capture, adiabatic transitions are the zeroth-order term in an
expansion in the defect-atom displacements that following capture (lattice
relaxation) and are, therefore, dominant under non-equilibrium conditions.
The electronic transition is caused instantaneously by the defect potential
(it is effectively a Franck-Condon transition) and the energy is dissipated
by MPP. The next order in the series, which is linear in the atomic
displacements, comprises two terms, only one of which has been captured by
prior theories \cite{Huang:1981,Alkauskas}. We estimate that these
\textquotedblleft linear terms\textquotedblright\ make smaller contributions
to the capture rate as they scale with $1/m$, where $m$ is a typical nuclear
mass in the defect complex. The adiabatic perturbation Hamiltonian that
couples the incoming electron to the defect is constructed in terms of
Hamiltonian matrices as in the Förster theory of electron and exciton
transfer in molecules \cite{MayBook}, which allows the derivation of Fermi's
golden rule for these transitions.

In addition to presenting the basic elements of the fundamental theory, we
report explicit calculations for capture cross sections as functions of
energy transfer for a prototype defect using DFT for the electronic matrix
elements. We employ the Projector-Augmented Wave (PAW)\ scheme \cite{BLOECHL}%
, which allows the use of the all-electron defect potential and wave
functions as opposed to pseudopotentials and pseudo wave functions. For the
calculation of the line-shape function, we introduce a Monte Carlo scheme to
sample the space of phonon combinations that contribute to the MPP energy
dissipation and find that random configurations containing up to \textit{%
twelve different phonon modes} and \textit{trillions} \textit{of
configurations} are needed to obtain converged results.

A few more observations are in order before we describe the present theory
in detail. In a perfect crystal without defects, the HRK perturbation
Hamiltonian is responsible for electron-phonon scattering (only linear
coupling is usually included) and for the formation of polarons, which are
electrons or holes dressed by phonons. Under strong-coupling conditions, the
HRK\ Hamiltonian can be responsible for polaron self-trapping. When a defect
is present, the HRK Hamiltonian can cause carrier capture. As Alkauskas et
al. \cite{Alkauskas} pointed out, such capture is very slow. Indeed it is
caused by the derivatives of the electronic wave functions with respect to
nuclear displacements, which amounts to a ``frozen electron approximation''
(recall that the BO approximation is effectively a ``frozen nuclei
approximation''). As we already noted, this kind of capture occurs under
thermal equilibrium conditions, which corresponds to constant emission and
capture by inverse processes, i.e., the band electrons are definitely
``aware'' of the defects, i.e., they should not be treated as ``free''
carriers with a mean free path, undergoing scattering by defects and
phonons. In this regard, the linear coupling approximation \cite{Alkauskas}
should be viewed as the zero mean-free-path limit, whereas the theory put
forward in this paper represents the limit in which the mean-free-path is
only bounded by $L_{\text{capture}}$, the mean distance an electron travels
before being captured by a defect.

The conditions under which capture cross sections are measured by junction
capacitance methods \cite{Henry:1977} are close to equilibrium, i.e., they
are slow. Similarly, in light-emitting diodes, carriers by design have
minimal acceleration through the pn-junction. However, even in such
deliberate setups, there must still be some nonequilibrium driving forces,
e.g., a current must flow through the system, in order to carry out the
measurement or for the device to operate. The carrier mean-free-path is
always finite, never exactly zero. % so that recombination at
%defects can be treated using the HRK\ Hamiltonian as in Ref. \cite{Alkauskas}
%(the linear term is probably adequate because it scales as $1/m$, whereas
%the quadratic term scales as $1/m^{2}$).  Here we are primarily interested
%in capture by hot electrons for which the HRK\ Hamiltonian is not
%sufficient. 
Therefore, a realistic model of the measured capture cross sections can be
obtained by scaling the difference between the two limits according to the
factor $L/L_{\text{capture}}$ where $L$ is the elastic scattering
mean-free-path, 
\begin{equation}
\sigma=\frac{L}{L_{\text{capture}}}\sigma_{\text{adiabatic}}+\sigma_{\text{%
nonadiabatic}},  \label{scaling}
\end{equation}
where $\sigma_{\text{nonadiabatic}}$ is the capture cross section due to the
HRK Hamiltonian and $\sigma_{\text{adiabatic}}$ is the adiabatic capture
cross section calculated in this paper.

For scattering of a carrier into another propagating state at a lower
energy, the defect is left in the same charge state, which requires that
scattering by the defect potential is elastic (no energy can be dissipated
in the Franck-Condon approximation in such a case). We find that inelastic
scattering can still occur within the BOA by the first-order correction to
the Franck-Condon approximation, which are the linear terms discussed above.

\section{Fermi golden rule for adiabatic and non-adiabatic transitions}

As discussed in the previous section, in order to describe transitions, it
is always necessary to identify the piece of the total Hamiltonian that
causes the transition between eigenstates of an approximate Hamiltonian. Let
us be more specific. In the hydrogen atom, one usually includes only the
Coulombic attraction between the proton and the electron, leaving out the
electromagnetic field at large. The calculated energy levels are only
eigenstates of this approximate Hamiltonian. The electromagnetic field,
treated as a perturbation, then causes a transition from, say, a \textit{2p}
state to the \textit{1s} state. In Auger transitions, one must leave out
specific electron-electron interactions that are then introduced to cause
transitions \cite{Laks:1990}. Our task here is to identify the approximate
Hamiltonian whose eigenstates are the propagating state of the incoming
electron that is not aware of the bound state of the defect potential and
the final state, which can be either another propagating state that is not
aware of the existence of a bound state at a lower energy or the bound state
itself, and determine the perturbation Hamiltonian that causes the
transition.

In the BOA, the many-electron Hamiltonian depends parametrically on the
nuclear positions and the total wave functions are products of many-electron
wave functions and phonon wave functions. Within DFT, the many-electron wave
functions are Slater determinants of Kohn-Sham wave functions. We start by
defining the many-electron Hamiltonian $H^{0}$ for the perfect crystal and
the corresponding eigenvalue problem, 
\begin{equation}
H^{0}|\Psi_{n}^{0}\rangle=E_{n}^{0}|\Psi_{n}^{0}\rangle.
\end{equation}
For the crystal containing a single defect, we have 
\begin{equation}
H|\Phi_{m}\rangle=E_{m}|\Phi_{m}\rangle.
\end{equation}
One normally writes 
\begin{equation}
H=H^{0}+\Delta H.  \label{HH}
\end{equation}

The partitioning of the total Hamiltonian $H$ according to Eq. (\ref{HH}) is
not useful for our purposes. Instead, we write 
\begin{equation}
H=\tilde{H}^{0}+H_{1}^{BO},  \label{HT}
\end{equation}
where, 
\begin{equation}
\tilde{H}^{0}|\Psi_{n}\rangle=\epsilon_{n}|\Psi_{n}\rangle.
\end{equation}
In order to obtain an explicit description of $H_{1}^{BO},$ which then
defines $\tilde{H}^{0}$ through Eq. (\ref{HT}), we express $\Delta H$ in
terms of the complete set of functions $\Psi_{n}:$ 
\begin{equation}
\Delta H=\sum_{m}|\Psi_{m}\rangle\langle\Psi_{m}|\Delta
H\sum_{n}|\Psi_{n}\rangle\langle\Psi_{n}|=\sum_{mn}|\Psi_{m}\rangle\Delta
H_{mn}\langle\Psi_{n}|.
\end{equation}
We then define $H_{1}^{BO}$ by 
\begin{equation}
H_{1}^{BO}=|\Psi_{i}\rangle\Delta
H_{if}\langle\Psi_{f}|+|\Psi_{f}\rangle\Delta H_{fi}\langle\Psi_{i}|,
\label{H1}
\end{equation}
where the subscripts $i$ and $f$ denote the eigenstates of $\tilde{H}^{0}$
that are the initial and final states of our problem. This definition of $%
H_{1}^{BO}$ is analogous to the so-called Förster transition often used in
energy transfer in molecules \cite{MayBook}. In effect, $H_{1}^{BO}$
eliminates the coupling of the incoming electron via the defect potential to
the final state, whether propagating or bound. The defect potential $\Delta
H,$ which can be arbitrarily strong, is still present. It is the
perturbation Hamiltonian $H_{1}^{BO}$ that is weak and can cause transitions
whose rate is describable by Fermi's gold rule, i.e., to first order in $%
H_{1}^{BO}.$ Note also that the state $|\Psi_{i}\rangle$ contains an
incoming electron that ``sees''\ the defect potential, but does not couple
to the bound state. Also, for all practical purposes, for carrier capture we
have $|\Psi_{f}\rangle=|\Phi_{f}\rangle$ (i.e., the bound sate is not
affected by the presence of an incoming electron that does not couple to the
defect).

The adiabatic transition rate is given by the usual Fermi's golden rule by 
\begin{equation}
w_{if}^{BO}=\frac{2\pi}{\hbar}\sum_{f}\left\vert \langle
X_{f}|\langle\Psi_{f}|H_{1}^{BO}|\Psi_{i}\rangle|X_{i}\rangle\right\vert
^{2}\delta(\Theta_{f}-\Theta_{i}+\epsilon_{if}),  \label{Fermi}
\end{equation}
where $\Theta_{i,f}$ are the total phonon energies of states $%
|X_{i,f}\rangle $ and $\epsilon_{if}=\epsilon_{f}-\epsilon_{i}$ is the
energy difference between the electronic states $|\Psi_{i}\rangle$ and $%
|\Psi_{f}\rangle$. For capture, it is usually assumed that there is one
final electronic state with a given energy difference $\epsilon_{if},$ but
there are many phonon configurations that can make up this difference. If
there are multiple electronic states at the same energy we need to sum Eq. (%
\ref{Fermi}) over all such states.

In addition to $H_{1}^{BO}$, there are terms beyond the BOA, usually
referred to as the non-adiabatic terms \cite{Huang:1950,Kubo:1952}, that
cause multiphonon transitions. These terms contain derivatives of the
electron wave functions with respect to nuclear coordinates $\{\mathbf{R}%
_{k}\}$ and are the terms neglected when one invokes the BOA. They
contribute to the total transition rate $w_{if}$ via the matrix element, 
\begin{equation}
-\sum_{k}\frac{\hbar^{2}}{2m_{k}}\left[\langle X_{f}|\langle\Psi_{f}|\nabla_{%
\mathbf{R}_{k}}^{2}\left(|\Psi_{i}\rangle|X_{i}\rangle\right)-\langle
X_{f}|\langle\Psi_{f}|\Psi_{i}\rangle\nabla_{\mathbf{R}_{k}}^{2}|X_{i}\rangle%
\right],  \label{Kubo}
\end{equation}
where $m_k$ is the mass of atom $k$. This contribution will be discussed in
detail later.

One can define a cross section for inelastic scattering or carrier capture
by 
\begin{equation}
\sigma_{if}=\frac{w_{if}\Omega}{v_{g}},
\end{equation}
where $v_{g}$ is the group velocity of the incident electron, $\Omega$ is
the volume over which the state $|i\rangle$ is normalized, so that $%
v_{g}/\Omega$ represents the flux of the incoming electrons.

We will work within DFT so that the many-electron wavefunctions are Slater
determinants of Kohn-Sham one-electron wavefunctions and the many-electron
Hamiltonians are those of non-interacting Kohn-Sham quasi-particles in the
presence of an effective single-particle external potential. From now on we
will view the Hamiltonians and wavefunctions in Eqs. (\ref{Fermi}) and (\ref%
{Kubo}) as one-electron Kohn-Sham Hamiltonians and electron wave functions
without change of notation.

\subsection{Adiabatic series}

We now examine the electronic part of the transition matrix element in the
BOA\ by showing explicitly its dependence on the atomic coordinates, 
\begin{equation}
M_{e}^{BO}(\{\mathbf{R}_{j}\})=\left\vert \langle\Psi_{f}(\{\mathbf{R}%
_{j}\})|H_{1}^{BO}(\{\mathbf{R}_{j}\})|\Psi_{i}(\{\mathbf{R}%
_{j}\})\rangle\right\vert ^{2}.
\end{equation}
The BOA by itself does not separate electron and phonon matrix elements. A
further approximation is needed. We expand 
\begin{equation}
M_{e}^{BO}(\{\mathbf{R}_{j}\})=M_{e}^{BO}(\{\mathbf{R}_{j}^{(0)}\})+\sum_{k}(%
\mathbf{R}_{k}-\mathbf{R}_{k}^{(0)})\cdot\nabla_{\mathbf{R}_{k}}M_{e}^{BO}(\{%
\mathbf{R}_{j}\})+\dots,
\end{equation}
in terms of the atomic displacements $\mathbf{R}_{k}-\mathbf{R}_{k}^{(0)}$
where $\mathbf{R}_{k}^{(0)}$ are the atomic positions in a reference state,
which will be determined later. The transition rate is then, 
\begin{eqnarray}
w_{if}^{BO} & = & \frac{2\pi}{\hbar}\left\vert M_{e}^{BO}(\{\mathbf{R}%
_{j}^{(0)}\})\right\vert ^{2}\sum_{f}\left\vert \langle
X_{f}|X_{i}\rangle\right\vert ^{2}\delta(\Theta_{f}-\Theta_{i}+\epsilon_{if})
\notag \\
& & +\frac{2\pi}{\hbar}\sum_{f}\left\vert \sum_{k}\nabla_{\mathbf{R}%
_{k}}M_{e}^{BO}(\{\mathbf{R}_{j}^{(0)}\})\cdot\langle X_{f}|(\mathbf{R}_{k}-%
\mathbf{R}_{k}^{(0)})|X_{i}\rangle\right\vert
^{2}\delta(\Theta_{f}-\Theta_{i}+\epsilon_{if})+....
\end{eqnarray}
Here the cross terms are dropped because the zeroth order and first order
terms cannot have the same final phonon wave functions -- the number of
phonons needed to ensure a nonzero overlap matrix element are different for
the two cases. The first term in this expansion represents a complete
separation of the electron and phonon wave functions as if they are
independent of each other and corresponds to the Frank-Condon approximation.
The second term is the first order correction to the Frank-Condon
approximation arising from the BOA perturbation Hamiltonian $H_{1}^{BO}$.

\subsection{Non-adiabatic series}

According to Huang \cite{Huang:1981}, the non-adiabatic matrix element
defined in Eq. (\ref{Kubo}) can be evaluated for linear phonon coupling, 
\begin{equation}
\sum_{k}\langle\Psi_{f}(\{\mathbf{R}_{j}^{(0)}\})|\nabla_{\mathbf{R}%
_{k}}H_{e}(\{\mathbf{R}_{j}^{(0)}\})|\Psi_{i}(\{\mathbf{R}%
_{j}^{(0)}\})\rangle\cdot\langle X_{f}|(\mathbf{R}_{k}-\mathbf{R}%
_{k}^{(0)})|X_{i}\rangle,  \label{linearPhonon}
\end{equation}
where $H_{e}$ is the electron part of the Hamiltonian. When electron-phonon
coupling $H_{ep}=H_{e}(\{\mathbf{R}_{j}\})-H_{e}(\{\mathbf{R}_{j}^{(0)}\})$
is introduced, the electron wave functions are changed by a perturbation, 
\begin{equation}
|\delta\Psi_{i}(\{\mathbf{R}_{j}\})\rangle=\sum_{i^{\prime}\neq i}\frac{%
\langle\Psi_{i^{\prime}}|H_{ep}|\Psi_{i}\rangle}{\epsilon_{i^{\prime}}-%
\epsilon_{i}}|\Psi_{i^{\prime}}(\{\mathbf{R}_{j}^{(0)}\})\rangle,
\label{deltapsi}
\end{equation}
(and a similar equation for the final states). We write both the initial and
final states in the form, 
\begin{equation}
|\Psi_{i(f)}(\{\mathbf{R}_{j}\})\rangle=|\Psi_{i(f)}(\{\mathbf{R}%
_{j}^{(0)}\})\rangle+|\delta\Psi_{i(f)}\rangle.
\end{equation}
Substituting this into Eq (\ref{Kubo}) and keeping only the linear terms, 
\begin{eqnarray}
& - & \sum_{k}\frac{\hbar^{2}}{2M_{k}}\left[\langle X_{f}|\nabla_{\mathbf{%
R_{k}}}^{2}\left(\langle\Psi_{f}(\{\mathbf{R}_{j}^{(0)}\})|\delta\Psi_{i}%
\rangle|X_{i}\rangle\right)-\langle X_{f}|\langle\Psi_{f}(\{\mathbf{R}%
_{j}^{(0)}\})|\delta\Psi_{i}\rangle\nabla_{\mathbf{R_{k}}}^{2}|X_{i}\rangle%
\right]  \notag \\
& = & \left(\Theta_{i}-\Theta_{f}\right)\langle X_{f}|\langle\Psi_{f}(\{%
\mathbf{R}_{j}^{(0)}\})|\delta\Psi_{i}\rangle|X_{i}\rangle  \notag \\
& = & \sum_{k}\sum_{i^{\prime}\neq i}\frac{\epsilon_{if}}{%
\epsilon_{i^{\prime}}-\epsilon_{i}}\langle\Psi_{i^{\prime}}|\nabla_{\mathbf{R%
}_{k}}H|\Psi_{i}\rangle\langle\Psi_{f}(\{\mathbf{R}_{j}^{(0)}\})|\Psi_{i^{%
\prime}}(\{\mathbf{R}_{j}^{(0)}\})\rangle\cdot\langle X_{f}|(\mathbf{R}_{k}-%
\mathbf{R}_{k}^{(0)})|X_{i}\rangle  \notag \\
& = & \sum_{k}\langle\Psi_{f}|\nabla_{\mathbf{R}_{k}}H_{e}|\Psi_{i}\rangle%
\cdot\langle X_{f}|(\mathbf{R}_{k}-\mathbf{R}_{k}^{(0)})|X_{i}\rangle.
\end{eqnarray}
Here the first equality results from the Schrödinger equations for the
phonon wave functions and for the second equality we used $%
\Theta_{i}-\Theta_{f}=\epsilon_{if}$.

We note that the above linear-order term in the non-adiabatic series has the
same phonon matrix element as the linear-order term in the BOA series of the
previous section. This indicates that the leading non-adiabatic term is a
smaller contribution to the electron capture rate compared to the
zeroth-order BOA term. The electronic matrix element in the non-adiabatic
series is different than the BOA series. We will show later that both these
terms scale as $1/m$, where $m$ is the mass of a typical atom in the defect
complex.

The linear term in Eq. (\ref{linearPhonon}) is usually referred to as the
linear electron-phonon coupling term. A similar term has been calculated by
Alkauskas et al \cite{Alkauskas}, with the exception that in that work the
wave functions are $|\Phi_{i(f)}\rangle$ which are the eigenstates of the
full Hamiltonian $H_{e}$, whereas in our case the wave functions are $%
\Psi_{i(f)}$ which are the eigenstates of the Hamiltonian $\tilde{H}^{0}$.
We recover the term calculated by Alkauskas et al. if we combine the BOA\
and the non-adiabatic series. We make use of the result in Eq. (\ref{gradMe}%
) and get for our final result

\begin{eqnarray}
w_{if} & = & \frac{2\pi}{\hbar}\left\vert M_{e}^{BO}(\{\mathbf{R}%
_{j}^{(0)}\})\right\vert ^{2}\sum_{f}\left\vert \langle
X_{f}|X_{i}\rangle\right\vert ^{2}\delta(\Theta_{f}-\Theta_{i}+\epsilon_{if})
\notag \\
& + & \frac{2\pi}{\hbar}\sum_{f}\left\vert \sum_{k}\left[\langle\Phi_{f}|%
\nabla_{\mathbf{R}_{k}}H_{e}|\Phi_{i}\rangle-\langle\Phi_{f}|\Psi_{i}^{0}%
\rangle\langle\Phi_{f}|\nabla_{\mathbf{R}_{k}}H_{e}|\Phi_{f}\rangle\right]%
\cdot\langle X_{f}|(\mathbf{R}_{k}-\mathbf{R}_{k}^{(0)})|X_{i}\rangle\right%
\vert ^{2}\times  \notag \\
& \times & \delta(\Theta_{f}-\Theta_{i}+\epsilon_{if})+\dots.
\label{Pseries}
\end{eqnarray}
Here the first term is the zeroth-rder term that corresponds to the
Franck-Condon approximation and thes second terms is the totality of
contributions from the linear terms in the two series. The first term in
square brackets is precisely the term that Alkauskas et al. \cite{Alkauskas}
calculated. We note that there exists a second term, which has the
appearance of a force term. These two terms can either add or subtract. We
will show shortly that these linear-order terms are proportional to $1/m$,
where $m$ is a typical atomic mass in the defect complex, and are,
therefore, significantly smaller than the zeroth-order Franck-Condon term,
which is dominant.

\section{Electron matrix elements}

We first consider the zeroth order term in the BOA series, which yields a
capture cross section that can be written in the familiar factorized form, 
\begin{equation}
\sigma_{if}=A_{if}F_{if},  \label{eq:AF}
\end{equation}
where $A_{if}$ contains the electronic part of the matrix element , 
\begin{equation}
A_{if}=\frac{\Omega}{\hbar v_{g}}\left|\langle\Psi_{f}(\{\mathbf{R}%
_{j}^{(0)}\})|H_{1}^{BO}(\{\mathbf{R}_{j}^{(0)}\})|\Psi_{i}(\{\mathbf{R}%
_{j}^{(0)}\})\rangle\right|^{2},
\end{equation}
and $F$ is called the line shape factor due to vibrations, 
\begin{equation}
F_{if}=\sum_{f}\left|\langle
X_{f}|X_{i}\rangle\right|^{2}\delta\left(\Theta_{f}-\Theta_{i}+\epsilon_{if}%
\right).  \label{eq:LSFgeneral}
\end{equation}
Next we will consider these two factors separately.

Detailed derivations given in Appendices \ref{A1} and \ref{A2} find the
final results 
\begin{equation}
M_{e}^{BO}=-\langle\Phi_{f}|\Psi_{i}^{0}\rangle\epsilon_{if},  \label{MeBO}
\end{equation}
and

\begin{equation}
\nabla_{\mathbf{R}_{k}}M_{e}^{BO}+\langle\Psi_{f}|\nabla_{\mathbf{R}%
_{k}}H|\Psi_{i}\rangle=\langle\Phi_{f}|\nabla_{\mathbf{R}_{k}}H|\Phi_{i}%
\rangle-\langle\Phi_{f}|\Psi_{i}^{0}\rangle\langle\Phi_{f}|\nabla_{\mathbf{R}%
_{k}}H|\Phi_{f}\rangle.  \label{gradMe}
\end{equation}

For the evaluation of the above matrix elements, we employ the PAW\ scheme,
which allows us to use all-electron wave functions instead of pseudo wave
functions. Details are given in Appendix \ref{A3}.

\section{Phonon matrix elements}

First, we consider the effect of displacements for a classical Hamiltonian.
We derive this Hamiltonian for the ion motion from which the phonon wave
functions and matrix elements can be calculated. For this purpose we start
with a supercell containing $n_{a}$ number of atoms with the defect site at
its center. This supercell is repeated $N$ times using the Born-von-Karman
periodic boundary condition. For the initial state, the equilibrium
positions of the atoms are $R_{k}$ where the subscript $k$ runs through both
the atomic index within the supercell and the cartesian components. Each
atom oscillates around its equilibrium position with displacement $u_{kl}$,
where the subscript $l$ labels different copies of the supercell under the
Born-von-Karman periodicity. Using the harmonic approximation for the
potential energy, under which only terms that are second order in
displacements make a contribution and introducing force constants $%
\Phi_{kl,k^{\prime}l^{\prime}}$, we can write \cite{MayBook}, 
\begin{equation}
H_{i}^{\prime}=\frac{1}{N}\sum_{kl}\left[\frac{1}{2}m_{k}\left(\frac{du_{kl}%
}{dt}\right)^{2}+\frac{1}{2N}\sum_{k^{\prime}l^{\prime}}u_{kl}\Phi_{kl,k^{%
\prime}l^{\prime}}u_{k^{\prime}l^{\prime}}\right]
\end{equation}
where the atomic mass $m_{k}$ also carries the subscript $k$ for convenience
even though it depends only on the atomic index and not the coordinate
component index.

When an electron is absorbed or emitted from the lattice, the equilibrium
position of the atoms change. The new equilibrium positions are $%
R_{k}+\Delta_{k}$. The new Hamiltonian has the same form after initial
displacement vectors $u_{kl}$ are replaced by $u_{kl}^{\prime}=u_{kl}-%
\Delta_{k}$. The final state Hamiltonian is then written as 
\begin{equation}
H_{f}^{\prime}=\frac{1}{N}\sum_{kl}\left\{ \frac{1}{2}m_{k}\left[\frac{%
d\left(u_{kl}-\Delta_{k}\right)}{dt}\right]^{2}+\frac{1}{2N}%
\sum_{k^{\prime}l^{\prime}}\left(u_{kl}-\Delta_{k}\right)\Phi_{kl,k^{%
\prime}l^{\prime}}\left(u_{k^{\prime}l^{\prime}}-\Delta_{k^{\prime}}\right)%
\right\}
\end{equation}
where we make an assumption that force constants do not change due to the
electron capture or absorption. Since displacements $\Delta_{k}$ do not
depend on time, the kinetic energy term remains unchanged. Expanding the
potential energy to first order in displacements reproduces the same term in
the original Hamiltonian plus a term that includes $u_{k^{\prime}l^{\prime}}%
\Delta_{k}$. 
\begin{equation}
H_{f}^{\prime}=H_{i}^{\prime}-\frac{1}{N}\sum_{kl,k^{\prime}l^{\prime}}%
\Phi_{kl,k^{\prime}l^{\prime}}\Delta_{k}u_{k^{\prime}l^{\prime}}
\end{equation}
Transforming to the normal-mode representation in terms of the generalized
coordinates, 
\begin{equation}
q_{j}=\frac{1}{\sqrt{N}}\sum_{kl}\sqrt{m_k}u_{kl}w_{j,kl},
\end{equation}
where $w_{j,kl}$ is the $kl$th element of the eigenvector for mode $j$. Note
that in this definition of the generalized coordinate $q_{j}$, it has
absorbed the mass factor $\sqrt{m_{k}}$. The Hamiltonian is expressed as, 
\begin{equation}
H_{f}^{\prime}=\frac{1}{2}\sum_{j}\dot{q}_{j}^{2}+\frac{1}{2}%
\sum_{j}\omega_{j}^{2}q_{j}^{2}-\frac{1}{\sqrt{N}}\sum_{j}q_{j}\sum_{kk^{%
\prime }}D_{kk^{\prime }}(\mathbf{k}_{j})w_{jk^{\prime }}\sqrt{m_k}%
\Delta_{k},  \label{eq:HamilonialDelta}
\end{equation}
where $\omega_{j}$ are the eigenfrequencies. A phase factor of the form $%
\exp(i\mathbf{k}_{j}\cdot\mathbf{r}_{l^{\prime}})$, where $\mathbf{k}_{j}$
is the wave vector of mode $j$, from $w_{j,k^{\prime }l^{\prime }}$ is
absorbed into the force constant matrix $\Phi$ yielding the dynamical matrix 
$D$, and reducing $w_{j,k^{\prime }l^{\prime }}$ to $w_{jk^{\prime }}$
(independent of $l^{\prime }$). Since we assume that force constants remain
the same after electron capture, 
\begin{equation}
\sum_{k^{\prime}}D_{kk^{\prime}}(\mathbf{k}_{j})w_{jk^{\prime}}=%
\omega_{j}^{2}w_{jk}
\end{equation}
The linear term causes a general coordinate displacement, 
\begin{equation}
\delta q_{j}=-\frac{1}{\sqrt{N}}\sum_{k}\sqrt{m_k}\Delta_{k}w_{jk}.
\end{equation}
We can express the normal coordinates of the lattice for the final ($f$)
state, $q_{j}^{f}$, in terms of those for the initial ($i$) state, $q_{j}$, 
\begin{equation}
q_{f,j}=q_{j}+\delta q_{j},
\end{equation}
so that the final Hamiltonian is:

\begin{equation}
H_{f}^{\prime}=\frac{1}{2}\sum_{j}\dot{q}_{f,j}^{2}+\frac{1}{2}%
\sum_{j}\omega_{j}^{2}q_{f,j}^{2}
\end{equation}

\subsection{Zeroth-order phonon matrix elements}

We have derived the expression for the generalized coordinates resulting
from the lattice displacements. These generalized displacements enter the
phonon wave functions $|X_{n_{j}^{i}}(q_{j})\rangle$ and $%
|X_{n_{j}^{f}}(q_{j}+\delta q_{j})\rangle$, respectively in the quantized
versions of the harmonic oscillator Hamiltonians $H_{i}^{\prime}$ and $%
H_{f}^{\prime}$ . Now we turn to the evaluation of phonon matrix elements $%
\langle X_{n_{j}^{f}}(q_{j}+\delta q_{j})|X_{n_{j}^{i}}(q_{j})\rangle$. When
the displacement $\delta q_{j}$ are small, we can show that the dominant
contribution comes from single phonon emission or absorption for each normal
mode. Suppose the initial state of mode $j$ has $n$ phonons and its final
state has $n+p$ phonons, (we dropped the index for the mode, since it is
present in the notation of generalized coordinate). Using the integrals
provided in Appendix \ref{A4}, the matrix elements for the phonon part are, 
\begin{equation}
\langle X_{n+1}(q_{j}+\delta q_{j})|X_{n}(q_{j})\rangle=-\sqrt{\frac{%
(n+1)\omega_{j}}{2\hbar}}\delta q_{j},
\end{equation}
\begin{equation}
\langle X_{n-1}(q_{j}+\delta q_{j})|X_{n}(q_{j})\rangle=\sqrt{\frac{%
n\omega_{j}}{2\hbar}}\delta q_{j}.
\end{equation}
The integrals for phonon modes that maintain the same occupation numbers are
calculated to second order in $q_{j}$, 
\begin{equation}
\langle X_{n}(q_{j}+\delta q_{j})|X_{n}(q_{j})\rangle=1-\frac{%
(2n+1)\omega_{j}}{4\hbar}\delta q_{j}^{2}.
\end{equation}

Now we consider how to evaluate Eq. (\ref{eq:LSFgeneral}). The total number
of phonon modes in the supercell is $M=3(n_{a}-1)$ excluding the
translational motion, and the total number of phonon modes in the entire
system is $MN$, since supercell is repeated $N$-times. We assume that there
is a one-to-one correspondence between phonon bands before and after the
capture. The wave function of the initial phonon state is 
\begin{equation}
\left\vert X_{i}\right\rangle =\prod_{j=1}^{MN}\left\vert
X_{n_{j}^{i}}\right\rangle ,  \label{eq:Xi}
\end{equation}
and that of any one of the final phonon states is 
\begin{equation}
\left\vert X_{f}\right\rangle =\prod_{j=1}^{MN}\left\vert
X_{n_{j}^{f}}\right\rangle ,  \label{eq:Xf}
\end{equation}
where $n_{j}^{i}$ and $n_{j}^{f}$ are the occupation numbers of phonon mode $%
j$ before and after the capture, and are also used to label the wave
functions. The total phonon energies for initial and final configurations
are 
\begin{equation}
\Theta_{i}=\frac{1}{N}\sum_{j=1}^{MN}n_{j}^{i}h\omega_{j}^{i},
\label{Thetai}
\end{equation}
and 
\begin{equation}
\Theta_{f}=\frac{1}{N}\sum_{j=1}^{MN}n_{j}^{f}h\omega_{j}^{f},
\label{Thetaf}
\end{equation}
respectively, where $\omega_{j}^{i}$ and $\omega_{j}^{f}$ is the phonon
frequency of mode $j$ in the initial and final configuration of the defect,
respectively. With the overlap matrix for each individual mode expressed as
Eq. (\ref{eq:overlap}) and using Eqs. (\ref{eq:Xi}), (\ref{eq:Xf}), (\ref%
{Thetai}), and (\ref{Thetaf}), Eq. (\ref{eq:LSFgeneral}) now takes the form, 
\begin{equation}
F_{if}=\sum_{\{n_{j}^{f}\}}\left\{ \prod_{j=1}^{MN}\left\vert \int
X_{n_{j}^{f}}(q_{j}+\delta q_{j})X_{n_{j}^{i}}(q_{j})dq_{j}\right\vert
^{2}\right\} \delta\left(\frac{1}{N}\sum_{j=1}^{MN}(n_{j}^{f}\hbar%
\omega_{j}^{f}-n_{j}^{i}\hbar\omega_{j}^{i})+\epsilon_{if}\right),
\label{eq:LSFproduct}
\end{equation}
where $n_{j}^{f}=n_{j}^{i}-1,n_{j}^{i},n_{j}^{i}+1$. We will see below that
as the limit of $N\rightarrow\infty$ is taken, the discrete modes in $N$
will become continuous spectra in $\mathbf{k}$ over the Brillouin zone of
the reciprocal space.

Now we are ready to put all the phonon matrix elements together and perform
the configurational sum. To do this we follow the steps of Huang and Rhys 
\cite{Huang:1950}, but generalize it for a system with multiple phonon
frequencies. For multiple phonon bands, we assume that the frequency
variation within each band is much smaller than the frequency difference
between the bands. This is the flat band approximation that is complemented
with the requirement of finite spacing between the bands. We finally find, 
\begin{equation}
F_{j}=\exp\left[\frac{p_{j}\hbar\omega_{j}}{2kT}-S_{j}\coth\left(\frac{%
\hbar\omega_{j}}{2kT}\right)\right]I_{p_{j}}\left[\frac{S_{j}}{%
\sinh(\hbar\omega_{j}/2kT)}\right],  \label{Fs}
\end{equation}
and 
\begin{equation}
F=\frac{1}{\Omega_{\mathbf{k}}}\sum_{\{p_{j}\}}\left.\left\{
\left(\prod_{j=1}^{M}F_{j}\right)\sum_{j=1}^{M}\left\{ p_{j}+\frac{S_{j}}{%
\sinh(\hbar\omega_{j}/2kT)}\frac{{\displaystyle I_{p_{j}+1}\left[\frac{S_{j}%
}{\sinh(\hbar\omega_{j}/2kT)}\right]}}{{\displaystyle I_{p_{j}}\left[\frac{%
S_{j}}{\sinh(\hbar\omega_{j}/2kT)}\right]}}\right\} D(\omega_{j})\right\}
\right\vert _{\sum_{j=1}^{M}p_{j}\hbar\omega_{j}+\epsilon_{if}=0},  \label{F}
\end{equation}
where 
\begin{equation}
S_{j}=\frac{\omega_{j}}{2\hbar}N\delta q_{j}^{2},
\end{equation}
and $I_{p}$ is the modified Bessel function of order $p$.

\subsection{Linear phonon matrix elements}

To evaluate the phonon matrix elements for the linear term, we rewrite it in
terms of the normal mode coordinates $q_{j}$, 
\begin{eqnarray}
\sum_{f}\left|\sum_{j}M_{j}\langle X_{f}|q_{j}|X_{i}\rangle\right|^{2} & = &
\sum_{f}\sum_{j}\left|M_{j}\langle X_{f}|q_{j}|X_{i}\rangle\right|^{2} 
\notag \\
& = & \frac{1}{2}\sum_{f}\left.\frac{\partial^{2}}{\partial\lambda^{2}}%
\left|\langle X_{f}|\prod_{j}\left[1+\lambda M_{j}q_{j}\exp(i\phi_{j})\right]%
|X_{i}\rangle\right|^{2}\right|_{\lambda=0},
\end{eqnarray}
where $\phi_{j}$ is a random phase introduced to cancel out the cross terms,
and, 
\begin{equation}
M_{j}=\langle\Phi_{f}|\partial_{q_{j}}H_{e}|\Phi_{i}\rangle-\langle\Phi_{f}|%
\Psi_{i}\rangle\langle\Psi_{f}|\partial_{q_{j}}H_{e}|\Psi_{f}\rangle.
\end{equation}
The rest of the steps are exactly the same as for the zeroth order matrix
elements. Using the integrals provided in Appendix \ref{A4}, the matrix
elements for the phonon part are, 
\begin{equation}
\langle X_{n+1}(q_{j}+\delta q_{j})|\left[1+\lambda M_{j}q_{j}\exp(i\phi_{j})%
\right]|X_{n}(q_{j})\rangle=-\sqrt{\frac{(n+1)\omega_{j}}{2\hbar}}\left[%
\delta q_{j}-\frac{\lambda\hbar M_{j}}{\omega_{j}}\exp(i\phi_{j})\right],
\end{equation}
\begin{equation}
\langle X_{n-1}(q_{j}+\delta q_{j})|\left[1+\lambda M_{j}q_{j}\exp(i\phi_{j})%
\right]|X_{n}(q_{j})\rangle=\sqrt{\frac{n\omega_{j}}{2\hbar}}\left[\delta
q_{j}+\frac{\lambda\hbar M_{j}}{\omega_{j}}\exp(i\phi_{j})\right],
\end{equation}
and, 
\begin{eqnarray}
\langle X_{n}(q_{j}+\delta q_{j})|\left[1+\lambda M_{j}q_{j}\exp(i\phi_{j})%
\right]|X_{n}(q_{j})\rangle &=& 1-\frac{(2n+1)\omega_{j}}{4\hbar}\delta
q_{j}^{2}-\frac{1}{2}\lambda M_{j}\delta q_{j}\exp(i\phi_{j})  \notag \\
&=&1-\frac{S_j}{2N}-\frac{1}{2}\lambda M_{j}\delta q_{j}\exp(i\phi_{j}).
\end{eqnarray}
Define, 
\begin{eqnarray}
S_{\pm}(\lambda)&=&%
\begin{pmatrix}
n+1 \\ 
n%
\end{pmatrix}%
\frac{\omega_j}{2\hbar}N\left|\delta q_{j}\mp\frac{\lambda\hbar M_{j}}{%
\omega_{j}}\exp(i\phi_{j})\right|^{2}  \notag \\
&\approx& 
\begin{pmatrix}
n+1 \\ 
n%
\end{pmatrix}%
\frac{\omega_j}{2\hbar}N\delta q_{j}^2\left|\exp\left[\mp 2\frac{%
\lambda\hbar M_{j}}{\omega_{j}\delta q_{j}}\exp(i\phi_{j})\right]\right|.
\end{eqnarray}
The approximation in the second step is accurate to $\lambda^2$, with the
ocnsideration that terms such as $\lambda^2 \sin 2\phi_j$ and $\lambda^2\cos
2\phi_j$ drop out after the configurational average. Then, 
\begin{equation}
\sqrt{S_{+}S_{-}}\approx \sqrt{n(n+1)}S_{j},
\end{equation}
\begin{equation}
\frac{S_{+}(\lambda)}{S_{-}(\lambda)}\approx\frac{n+1}{n}\left|\exp\left[-4%
\frac{\lambda\hbar M_{j}}{\omega_{j}\delta q_{j}}\exp(i\phi_{j})\right]%
\right|.
\end{equation}
The $\lambda$-dependent line-shape factor for a single phonon band is, 
\begin{eqnarray}
F_{j}(\lambda) & = & \exp\left[\frac{p_{j}\hbar\omega_{j}}{2kT}%
-S_{j}\coth\left(\frac{\hbar\omega_{j}}{2kT}\right)-\lambda N \delta
q_{j}\left|M_{j}\exp(i\phi_{j})\right|\right]I_{p_{j}}\left[\frac{S_{j}}{%
\sinh(\hbar\omega_{j}/2kT)}\right]\times  \notag \\
& & \left|\exp\left[ -2\lambda p_{j}\frac{\hbar M_{j}}{\omega_{j}\delta q_{j}%
}\exp(i\phi_{j})\right]\right|.
\end{eqnarray}
Let us now compare the two $\lambda$ factors by evaluating the ratio 
\begin{equation}
\frac{N\omega_j\delta q_j^2}{2\hbar}=\frac{m\omega_j\delta R^2}{\hbar}.
\end{equation}
For a hydrogenated vacancy defect our calculation shows that $\delta
R\approx0.2$ Å\ for the nearest Si atom. Using $m\approx4.66\times10^{-26}$
kg for the Si atom and $\omega_j\approx 10^{12}$ sec$^{-1}$, we have, 
\begin{equation}
\frac{N\omega_j\delta q_j^2}{2\hbar}\approx 0.09.
\end{equation}
Thus the first $\lambda$ factor has a much smaller contribution than the
second one. The final linear phonon squared matrix element is, 
\begin{eqnarray}
F_{1} & = & \frac{1}{2\Omega_{\mathbf{k}}}\sum_{\{p_{j}\}}\left\{ \frac{%
\partial^{2}}{\partial\lambda^{2}}\left.\left[\prod_{j=1}^{M}F_{j}(\lambda)%
\right]\right|_{\lambda=0}\right.  \notag \\
& & \left.\left.\sum_{j=1}^{M}\left\{ p_{j}+\frac{S_{j}}{\sinh(\hbar%
\omega_{j}/2kT)}\frac{{\displaystyle I_{p_{j}+1}\left[\frac{S_{j}}{%
\sinh(\hbar\omega_{j}/2kT)}\right]}}{{\displaystyle I_{p_{j}}\left[\frac{%
S_{j}}{\sinh(\hbar\omega_{j}/2kT)}\right]}}\right\} D(\omega_{j})\right\}
\right|_{\sum_{j=1}^{M}p_{j}\hbar\omega_{j}+\epsilon_{if}=0}.  \label{F1}
\end{eqnarray}

\subsection{Ratio of zeroth-order and linear terms}

From the different expressions for the zeroth-order and the linear phonon
matrix elements, we can estimate the ratio between the linear term and the
zeroth-order term in the transition rate. This is of the order of 
\begin{equation}
2\left\vert \frac{M_{j}\hbar p_{j}}{M_{e}^{BO}\omega_{j}\delta q_{j}}%
\right\vert ^{2}.
\end{equation}
To estimate $M_{j}/M_{e}^{BO}$, we note that the leading term in $M_{j}$ is
(see Eq. (\ref{leading})), 
\begin{equation}
M_{j}\approx-\epsilon_{if}\langle\frac{\partial\Phi_{f}}{\partial q_{j}}%
|\Psi_{i}\rangle.
\end{equation}
To estimate $\partial\Phi_{f}/\partial q_{j}$, we assume rigid atomic
orbitals, where the atomic wave functions move rigidly in space with each
atom. The derivative of such a wave function with respect to atomic
displacements simply reflects the change in the relative spatial phase,
which is dictated by the phonon wave vector, 
\begin{equation}
\frac{\partial\Phi_{f}}{\partial q_{j}}\approx i\sqrt{\frac{N}{m}}\frac{2\pi%
}{\lambda_{j}}\Phi_{f}\exp(i\phi),
\end{equation}
where $\lambda_{j}$ is the acoustic wavelength for mode $j$, $m$ is the mass
of an atom, and $\phi$ is the phase factor due to the movement of the atoms,
which is different in each Born-von-Karman supercell. Integrating over all $%
N $ Born-von-Karman supercells, the sum of the $\exp(i\phi)$ factors scales
as $1/N$ for large $N$. Thus, 
\begin{equation}
M_{j}\approx i\frac{2\pi}{\sqrt{Nm}\lambda_{j}}M_{e}^{BO}.
\end{equation}
Finally, $p_j$ is mostly zero, occasionally taking the values $\pm 1$, and $%
\delta q_{j}\approx\sqrt{(m/N)}\delta R$ where $\delta R$ is the largest
atomic displacement and $m$ is the mass of the corresponding atom. The ratio
between the linear and zeroth order terms simplifies to, 
\begin{equation}
2\left(\frac{\hbar}{cm\delta R}\right)^{2},
\end{equation}
where $c$ is the sound velocity in the material. For a hydrogenated vacancy
defect our calculation shows that $\delta R\approx0.2$ Å\ for the nearest Si
atom. Using this number and $c\approx8\times10^{3}$ m$/$s for bulk silicon
and $m\approx4.66\times10^{-26}$ kg for the Si atom, we find, 
\begin{equation}
2\left(\frac{\hbar}{cm\delta R}\right)^{2}\approx 3.6\times 10^{-4}.
\end{equation}
Thus the linear phonon term (non-adiabatic term) is several orders of
magnitude smaller than the leading BOA term.

\subsection{Monte Carlo method for configurational sum}

The summation over all configurations $\{p_{j}\}$ involves a large number of
terms when $P=\sum_{j}|p_{j}|$ is greater than a few. We use a Monte Carlo
approach to calculate this sum. For a given number of phonon modes, $P$, and
a given number of bands, $B$, we use Monte-Carlo to construct a fixed number
of configurations, $K$. We rewrite the sum over the configurations as a sum
over the number of phonons $P$ of a configuration, a sum over the number of
bands $B$ used to construct a configuration with $P$ phonons and a sum over
the configurations sampled (Monte Carlo steps). In each Monte Carlo step, we
randomly pick $B$ bands and then we construct all the possible
configurations with $P$ phonons constructed by these bands.

In order to generate and count the configurations correctly, we first
rewrite Eq. (\ref{F}) as, 
\begin{eqnarray}
F & = & \frac{1}{\Omega_{\mathbf{k}}}\sum_{P=1}\sum_{B=1}^{P}w_{B}\sum_{%
\{p_{j}\}\prime}^{K}\left\{ \left(\prod_{j=1}^{M}F_{j}\right)\right.\times 
\notag \\
& & \left.\left.\sum_{j=1}^{M}\left\{ p_{j}+\frac{S_{j}}{\sinh(\hbar%
\omega_{j}/2kT)}\frac{{\displaystyle I_{p_{j}+1}\left[\frac{S_{j}}{%
\sinh(\hbar\omega_{j}/2kT)}\right]}}{{\displaystyle I_{p_{j}}\left[\frac{%
S_{j}}{\sinh(\hbar\omega_{j}/2kT)}\right]}}\right\} D(\omega_{j})\right\}
\right|_{\sum_{j=1}^{M}p_{j}\hbar\omega_{j}+\epsilon_{if}=0}.  \label{F'}
\end{eqnarray}

Then, we normalize the sum so that the total weight, $w_{B}$, in each
sub-group of configurations (configurations with the same number of bands)
is equal to the total number of possible configurations for this number of
bands, 
\begin{equation}
w_{B}=\frac{1}{K}\frac{M!}{B!(M-B)!}.  \label{weight}
\end{equation}

All configurations with up to four phonon modes are constructed and
calculated explicitly. For configurations with more than four phonons, all
the configurations constructed with up to three bands are calculated
explicitly and the above equations are use to calculate the line shape
function for configurations with more than three bands.

The last step in the Monte Carlo scheme is to collect the line shape
function into different energy bins for a distribution. To do this, we note
that with an incomplete sampling of the phase space via Monte Carlo, we may
not be able to resolve the energy distribution to arbitrary accuracy.
Specifically, when we sample one configuration and weigh it according to Eq.
(\ref{weight}), we are effectively using it to approximate several
configurations with different energies. Thus, the energy resolution must be
consistent with the number of configuration samples - fewer configurations
should correspond to coarser energy resolution. For this reason, we define
the energy bin width separately for each value of $P$ based on the
requirement that there is at least one configuration inside each energy bin.
To ensure the correct normalization, we rewrite the phonon density of states
for band $j$ as, 
\begin{equation}
D(\omega_{j})=\frac{1}{\Delta E}\int D(E)dE=\frac{\Omega_{\mathbf{k}}}{%
\Delta E}
\end{equation}
where $\Delta E$ is the energy bin width and we assume that the phonon band
is sufficiently flat so that it falls entirely within one energy bin. Then
Eq. (\ref{F'}) becomes, 
\begin{eqnarray}
F & = & \frac{1}{\Delta E}\sum_{P=1}\sum_{B=1}^{P}w_{B}\sum_{\{p_{j}\}%
\prime}^{K}\left\{ \left(\prod_{j=1}^{M}F_{j}\right)\right.\times  \notag \\
& & \left.\left.\sum_{j=1}^{M}\left\{ p_{j}+\frac{S_{j}}{\sinh(\hbar%
\omega_{j}/2kT)}\frac{{\displaystyle I_{p_{j}+1}\left[\frac{S_{j}}{%
\sinh(\hbar\omega_{j}/2kT)}\right]}}{{\displaystyle I_{p_{j}}\left[\frac{%
S_{j}}{\sinh(\hbar\omega_{j}/2kT)}\right]}}\right\} \right\} \right\vert
_{\sum_{j=1}^{M}p_{j}\hbar\omega_{j}+\epsilon_{if}=0}.
\end{eqnarray}

The evaluation of the linear phonon terms is similar.

\section{Application to a defect in silicon}

In this paper, we will present only one application of the theory and
computer codes for the capture cross section of a prototype defect in Si,
namely a triply hydrogenated vacancy with a bare dangling bond. Our purpose
here is to demonstrate the feasibility of calculations, especially the
first-ever calculation of the line-shape function that is converged with
respect to the number of phonon modes that are used to construct random
configurations whose energy is equal to the amount of energy that needs to
be dissipated following the instantaneous electronic transition. We defer
calculations for defects for which experimental data are available to a
future paper where we anticipate using hybrid functionals in the DFT
calculations of the electronic matrix elements. Such calculations are
computationally demanding, but would provide more accurate transition
energies and electronic matrix elements. In addition, we plan to code the
additional contributions from the linear terms which we estimated to be
significantly smaller because they scale with the inverse of the mass of a
typical atom in the defect cluster. It will be interesting to see how the
two terms in the square brackets in eq. \ref{Pseries} add or subtract for
different defects.

\begin{figure}[htbp]
\centering \includegraphics[width=1\textwidth]{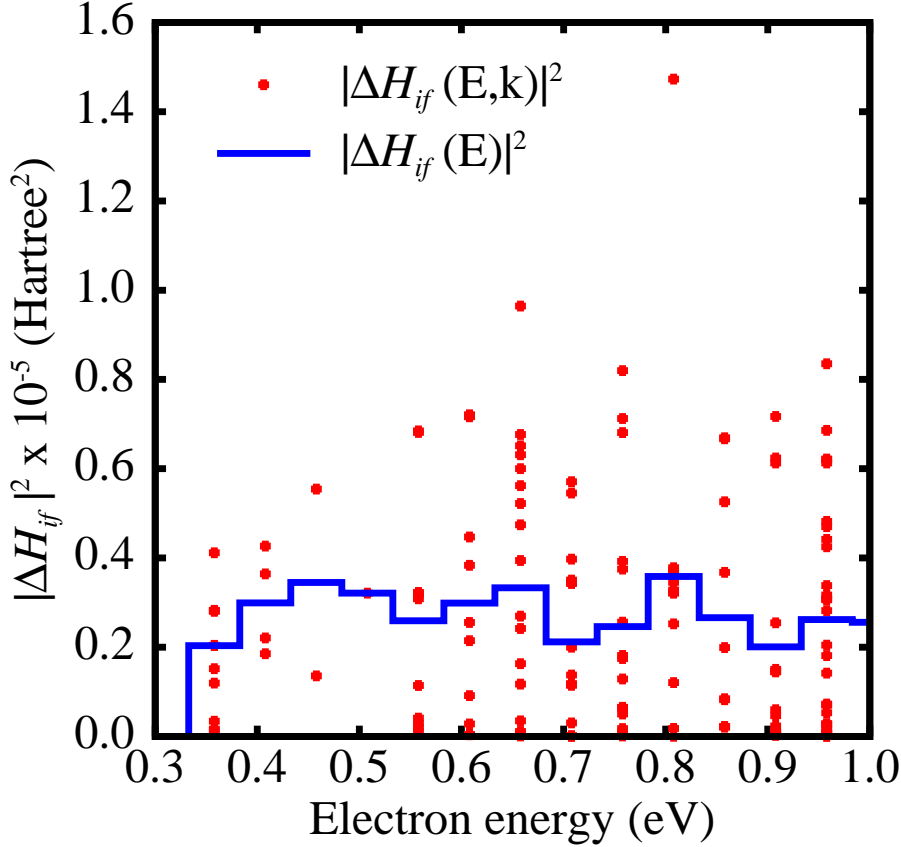}\newline
\caption{Calculated electronic matrix elements as a function of the initial
state electron energy for a triply hydrogenated vacancy in Si with a bare
dangling bond. Red points: matrix element values at each energy for
different $k$ points; blue curve: Averaged matrix element over all $k$
points for each energy.}
\label{TME}
\end{figure}

In Fig. 1, we show the values of calculated electronic matrix elements as a
function of energy. At each energy value, there are a number of $k$ points
that contribute. Their contributions are indicated by red symbols. The size
of the energy bin is determined by the number of $k$ points. For the example
shown in Fig. 1 the average matrix element as a function of energy is shown
by the blue line. The size of the energy bin fixes the resolution. A smooth
curve can only be obtained with very small energy bins, which requires a
very large number of $k$ points. It is clear from the figure that the
capture electronic matrix element is relatively constant as a function of
energy, whereby it seems best at this point to take it to be a constant,
either an average value or the value at the threshold for capture, which
introduces an error bar of a factor of $\sim1.7$ (clearly, to validate the
theory against accurate experimental data, we need a very accurate
calculation in the near-threshold region).

\begin{figure}[tbph]
\centering \includegraphics[width=1%
\textwidth]{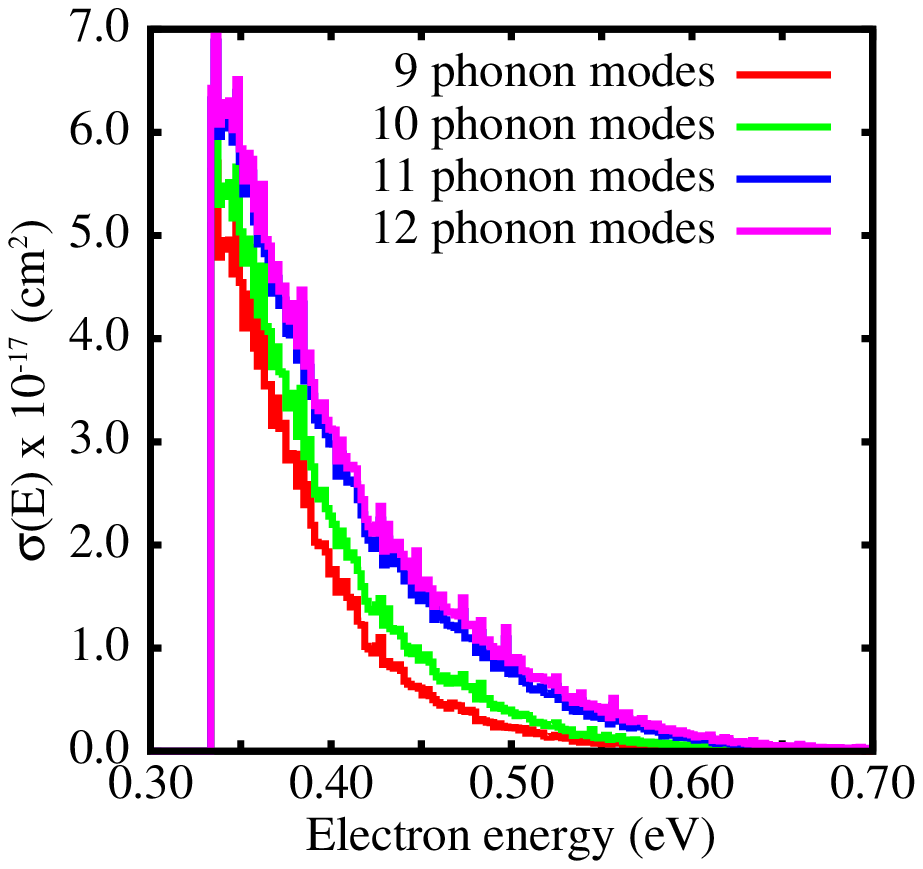}\newline
\caption{Calculated electron capture cross section using a constant electron
matrix element and different number of phonon modes.}
\label{SigmaConverge}
\end{figure}

In Fig. 2, we show the calculated capture cross section using a constant
matrix element to show clearly the convergence of the line-shape function as
we increase the number of phonon modes that are used to construct
configurations (the electronic matrix element is just a multiplier that sets
the absolute value). The dominant contribution to the line-shape function
comes from the balance between the modes with largest general coordinate
displacement (GCD) and the growth of the number of allowed combinations with
smaller GCD. Note that the curves are smooth because we employ millions of
configurations at each energy and therefore we have very tiny energy bins.
It is clear that a single-phonon-mode approximation would be very poor
indeed. In Fig. 3 we show the convergence of the capture cross section at
threshold (for electrons at the bottom of the conduction band), which is
what is usually measured. Once more, it is clear that the single-phonon mode
approximation would be inadequate. 
\begin{figure}[tbph]
\centering \includegraphics[width=1%
\textwidth]{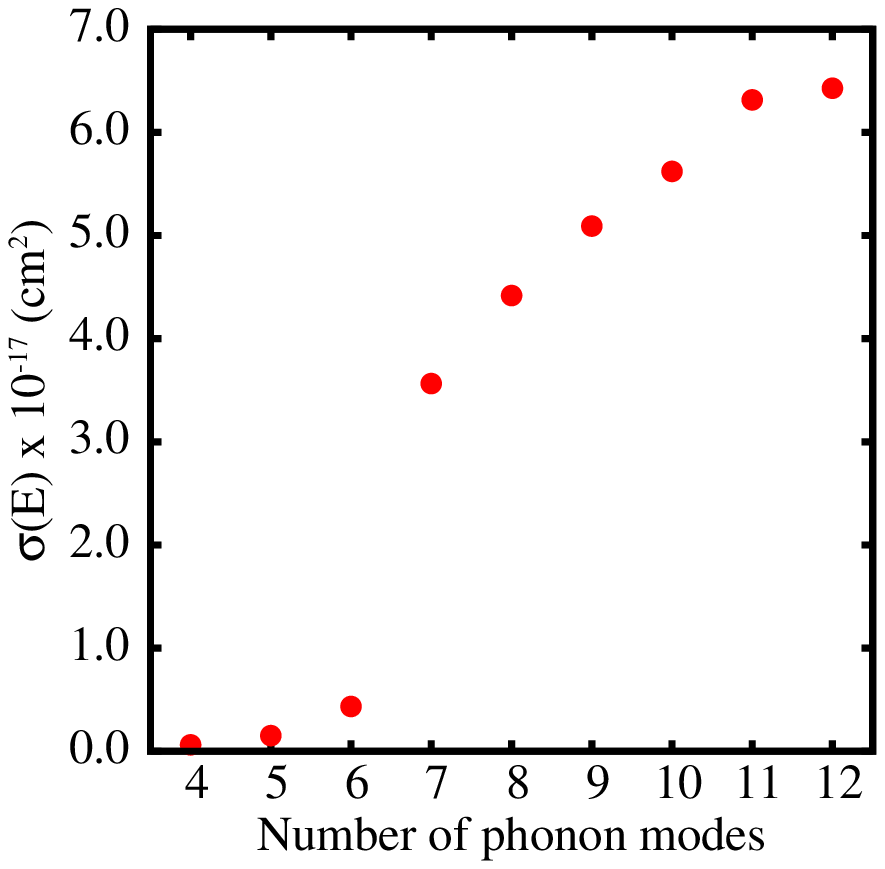}\newline
\caption{Convergence of the calculated electron capture cross section at the
threshold as a function of the number of phonon modes.}
\label{SigmaConverge}
\end{figure}

For a calculation of the cross section using electronic matrix elements that
depend on energy, the resolution is limited by the energy bin size. We show
the result in Fig. 4. Clearly, the size of the energy bin is important. For
capture cross sections, one is often interested only in the threshold value.
The calculations presented here are a prelude to calculations of
hot-electron inelastic multiphonon scattering, for which the energy
dependence is important. The energy dependence is also important in
luminescence curves, i.e., the classic Huang-Rhys problem that was treated
in the single-phonon approximation in Ref. \onlinecite{REF2012} (in the case
of luminescence, MPPs dissipate only the relaxation energy of the defect,
when one expects the phonon mode corresponding to the actual relaxation to
dominate; nevertheless, a fully convergent calculation would be needed to
establish the degree of accuracy one obtains with the single-mode
approximation). 
\begin{figure}[tbph]
\centering \includegraphics[width=1\textwidth]{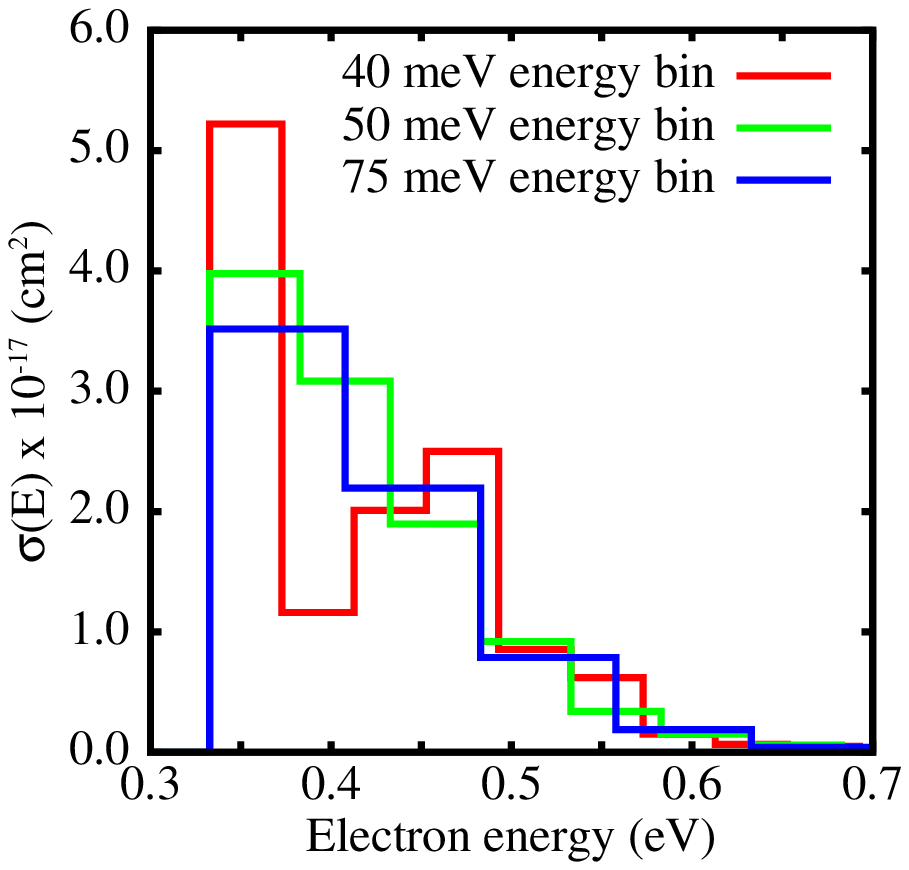}\newline
\caption{Calculated full capture cross section using the electron matrix
element from Fig. 1 and 12 phonon modes in the line-shape function.}
\label{Sigma}
\end{figure}
\begin{figure}[tbph]
\centering \includegraphics[width=1\textwidth]{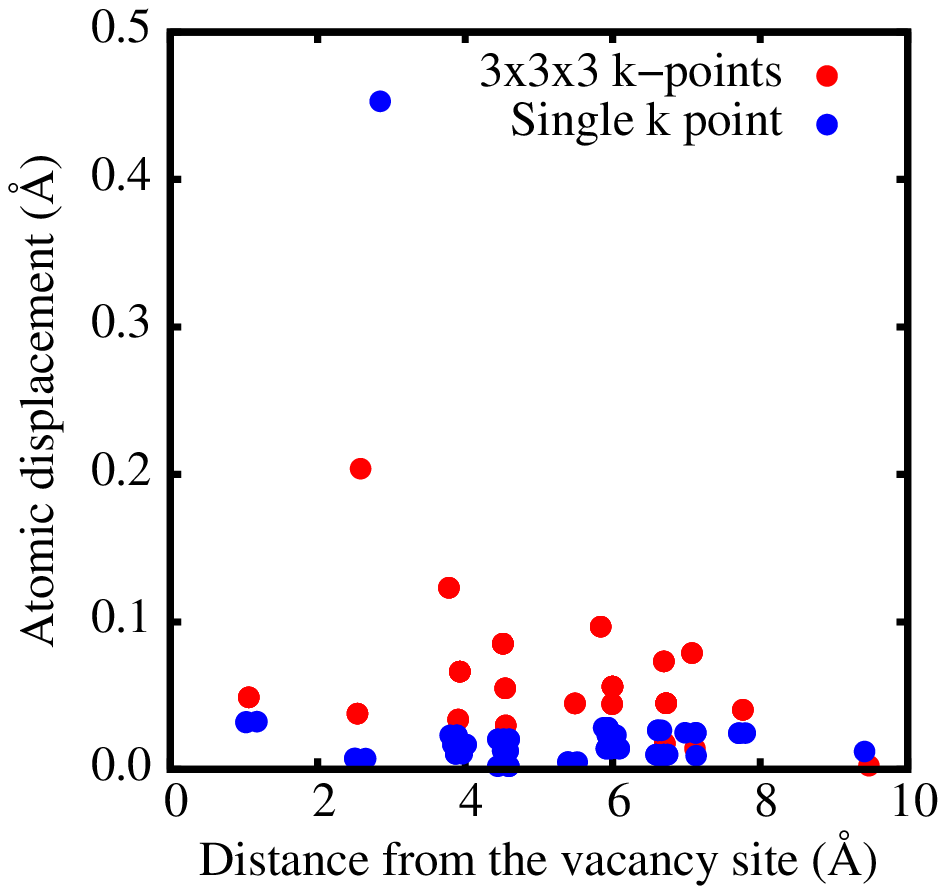}%
\newline
\caption{Atomic displacements of the triply-hydrogenated Si vacancy as a
function of the distance from the vacancy site for a 64-atom supercell.}
\label{displacements}
\end{figure}
The accuracy of the calculation of the line shape function is controlled by
the accuracy of the calculation of the generalized displacements. The latter
depends on the accuracy of the calculation of the atomic displacements. We
found that accuracy is enhanced significantly if we allow the entire
supercell to relax, which allows the defect's neighbors to relax more
freely. At the same time, a dense $k$-point mesh is necessary. In Fig. 5, we
present the atomic displacements of the triply-hydrogenated Si vacancy as a
function of the distance from the vacancy site for a 64-atom supercell.
Using only one $k$-point and not allowing the supercell to relax we get only
the Si-atom near the defect to move significantly while the rest of the
crystal remains essentially frozen (blue dots). This kind of relaxation
leads to only a few phonon modes being significant and thus the system is
artificially able to dissipate energy efficiently at certain frequencies. On
the other hand the well-relaxed crystal of the ($3\times 3\times 3$) $k$%
-points grid (red dots) has more atoms contributing to the generalized
displacements and thus almost all the phonon modes contribute in the
dissipation to the energy of the incoming electron. The use of supercells
with more than 64 atoms would be prohitively expensive for the line-shape
function calculation.

\section{Summary}

We have presented a comprehesive theory of inelastic multiphonon carrier
capture and scattering processes. We showed that, under non-equilibrium
conditions, i.e., in the presence of currents or hot electrons, the defect
potential is primarily responsible for capture throught a zeroth-order term
in an expansion in terms of the atomic displacements (relaxation)\ that
accompanies capture. These terms were not included in any prior theory.
Instead, the focus has always been on the linear terms, which we showed here
to be much smaller because they depend on the inverse of the mass of typical
atoms in the defect complex. The linear terms are dominant only in the limit
of thermal equilibrium. For the first time, we used accurate all-electron
wave functions obtained by the PAW\ method for the electronic matrix
elements and an accurate Monte Carlo scheme to sample random configurations
of up to 12 distinct phonon modes for the line-shape functions to achieve
convergence (a single-phonon-mode approximation has been standard in prior
calculations).\ We presented results for a prototype defect. More accurate
hybrid exchange-correlation functionals are needed to produce results that
are accurate enough for comparison with experimental data. In addition, a
reliable comparison with data can only be made with experimental
measurements of capture cross sections simultaneously with the determination
of the elastic mean-free-path and the capture mean-free-path, as they appear
in Eq. (\ref{scaling}).

\acknowledgements

We would like to thank Chris Van de Walle and Audrius Alkauskas for valuable
discussions. This work was supported in part by the Samsung Advanced
Institute of Technology (SAIT)'s Global Research Outreach (GRO) Program, by
the AFOSR and AFRL through the Hi-REV program, and by NSF\ grant
ECCS-1508898. \ A portion of this research was conducted at the Center for
Nanophase Materials Sciences, which is sponsored at Oak Ridge National
Laboratory by the Division of Scientific User Facilities. The computation
was done using the utilities of the National Energy Research Scientific
Computing Center (NERSC) and resources of the Oak Ridge Leadership Computing
Facility at the Oak Ridge National Laboratory, which is supported by the
Office of Science of the U.S. Department of Energy under Contract No.
DE-AC05-00OR22725. The work was also supported by the McMinn Endowment at
Vanderbilt University.

{\appendix
}

\section{Evaluation of the electronic matrix element for BOA transition}

\label{A1}

In the basis of $|\Psi_{n}\rangle$, the unperturbed Hamiltonian $\tilde{H}%
^{0}$ is diagonal with eigenenergies $\epsilon_{n}$. The total electron
Hamiltonian $H=\tilde{H}^{0}+H_{1}^{BO}$ has coupling terms only between
states $|\Psi_{i}\rangle$ and $|\Psi_{f}\rangle$. We can, therefore,
construct solutions of 
\begin{equation}
(\tilde{H}^{0}+H_{1}^{BO})|\Phi\rangle=E|\Phi\rangle.  \label{eg:H0Phi}
\end{equation}
in the form $|\Phi\rangle=a|\Psi_{i}\rangle+b|\Psi_{f}\rangle,$ so that 
\begin{equation}
\begin{pmatrix}
\epsilon_{i} & \Delta H_{if} \\ 
\Delta H_{fi} & \epsilon_{f}%
\end{pmatrix}%
\begin{pmatrix}
a \\ 
b%
\end{pmatrix}%
=E%
\begin{pmatrix}
a \\ 
b%
\end{pmatrix}%
.
\end{equation}
There are two sets of solutions, 
\begin{equation}
E_{i(f)}=\frac{1}{2}\left[\epsilon_{i}+\epsilon_{f}\pm\sqrt{%
(\epsilon_{i}-\epsilon_{f})^{2}+4|\Delta H_{if}|^{2}}\right],
\label{epsilon_if}
\end{equation}
where state $i$ takes the $+$ sign and state $f$ take the $-$ sign, since $%
E_{i}>E_{f}$. The coefficients satisfy 
\begin{equation}
\epsilon_{i}a_{i}+\Delta H_{if}b_{i}=E_{i}a_{i}
\end{equation}
and 
\begin{equation}
|a_{i}|^{2}+|b_{i}|^{2}=1.
\end{equation}
There is an arbitrary phase factor within $a_{i}$. We can define a set of
solution as, 
\begin{equation}
a_{i}=b_{f}^{\ast}=\sqrt{\frac{1}{2}+\sqrt{\frac{1}{4}-\left\vert \frac{%
\Delta H_{if}}{E_{i}-E_{f}}\right\vert ^{2}}}
\end{equation}
and 
\begin{equation}
b_{i}=-a_{f}^{\ast}=\frac{\Delta H_{if}}{E_{i}-E_{f}}\frac{1}{{\displaystyle 
\sqrt{\frac{1}{2}+\sqrt{\frac{1}{4}-\left\vert \frac{\Delta H_{if}}{%
E_{i}-E_{f}}\right\vert ^{2}}}}}.  \label{eq:bi}
\end{equation}

If we can compute the overlap integral $\langle\Phi_{f}|\Psi_{i}%
\rangle=a_{f} $, then we can solve for $|\Delta H_{if}|^{2}$ from $%
|a_{f}|^{2}$ and find, 
\begin{equation}
|\Delta H_{if}|^{2}=\frac{|\langle\Phi_{f}|\Psi_{i}\rangle|^{2}-|\langle%
\Phi_{f}|\Psi_{i}\rangle|^{4}}{\left(1-2|\langle\Phi_{f}|\Psi_{i}%
\rangle|^{2}\right)^{2}}\epsilon_{if}^{2}.
\end{equation}
To be consistent with the phase of Eq. (\ref{eq:bi}), we have, 
\begin{equation}
M_{e}^{BO}=\langle\Psi_{f}|H_{1}^{BO}|\Psi_{i}\rangle=\Delta H_{if}=-\frac{%
\sqrt{1-|\langle\Phi_{f}|\Psi_{i}\rangle|^{2}}}{1-2|\langle\Phi_{f}|\Psi_{i}%
\rangle|^{2}}\langle\Phi_{f}|\Psi_{i}\rangle\epsilon_{if}.
\end{equation}
The wave function $|\Psi_{i}\rangle$ is related to that of a perfect crystal 
$|\Psi_{i}^{(0)}\rangle$ through a perturbation expansion, 
\begin{equation}
|\Psi_{i}^{(0)}\rangle=|\Psi_{i}\rangle-\sum_{i^{\prime}\neq i,f}\frac{%
\langle\Psi_{i^{\prime}}|\Delta H|\Psi_{i}\rangle}{\epsilon_{i^{\prime}}-%
\epsilon_{i}}|\Psi_{i^{\prime}}\rangle.
\end{equation}
Because $H_{1}$ has only nonzero elements between the states $%
|\Psi_{i}\rangle$ and $|\Psi_{f}\rangle$, for $j\neq i,f$, the wave
functions $|\Psi_{j}\rangle=|\Phi_{j}\rangle$ so that $\langle\Phi_{f}|%
\Psi_{j}\rangle=0$. Thus, to first order in the defect potential, 
\begin{equation}
\langle\Phi_{f}|\Psi_{i}\rangle=\langle\Phi_{f}|\Psi_{i}^{0}\rangle,
\label{FP}
\end{equation}
and, assuming that $|\langle\Phi_{f}|\Psi_{i}^{0}\rangle|\ll1$, we arrive at
Eq. (\ref{MeBO}), which simplifies the evaluation of the overlap integral.

\section{Evaluation of the gradient terms}

\label{A2}

Using the result in the previous section for the matrix element $M_{e}^{BO}$%
, we now calculate the gradient terms in Eq. (\ref{Pseries}), $\nabla_{%
\mathbf{R}_{k}}M_{e}^{BO}+\langle\Psi_{f}|\nabla_{\mathbf{R}%
_{k}}H_{e}|\Psi_{i}\rangle$. Neglecting higher order $|\langle\Phi_{f}|%
\Psi_{i}\rangle|^{2}$ terms, the first gradient term is, 
\begin{eqnarray}
\nabla_{\mathbf{R}_{k}}M_{e}^{BO} & = & -\left(\langle\nabla_{\mathbf{R}%
_{k}}\Phi_{f}|\Psi_{i}\rangle+\langle\Phi_{f}|\nabla_{\mathbf{R}%
_{k}}\Psi_{i}\rangle\right)\epsilon_{if}-\langle\Phi_{f}|\Psi_{i}\rangle%
\nabla_{\mathbf{R}_{k}}\epsilon_{if}  \notag \\
& = & -\left(\langle\nabla_{\mathbf{R}_{k}}\Phi_{f}|\Psi_{i}\rangle+\langle%
\Phi_{f}|\nabla_{\mathbf{R}_{k}}\Psi_{i}\rangle\right)\epsilon_{if}-\langle%
\Phi_{f}|\Psi_{i}\rangle\langle\Psi_{f}|\nabla_{\mathbf{R}%
_{k}}H_{0}|\Psi_{f}\rangle,
\end{eqnarray}
where in the last step we used the fact that $\nabla_{\mathbf{R}%
_{k}}\epsilon_{i}=0$ (the initial state is at equilibrium) and the
Helmann-Feynman theorem for $\nabla_{\mathbf{R}_{k}}\epsilon_{f}$. From Eq. (%
\ref{deltapsi}) we have, 
\begin{equation}
|\nabla_{\mathbf{R}_{k}}\Psi_{i}\rangle=\sum_{i^{\prime}\neq i}\frac{%
\langle\Psi_{i^{\prime}}|\nabla_{\mathbf{R}_{k}}H_{e}|\Psi_{i}\rangle}{%
\epsilon_{i^{\prime}}-\epsilon_{i}}|\Psi_{i^{\prime}}\rangle,
\end{equation}
where we used $\nabla_{\mathbf{R}_{k}}H_{el}=\nabla_{\mathbf{R}_{k}}H_{e}$.
Because $|\Psi_{i^{\prime}}\rangle=|\Phi_{i^{\prime}}\rangle$ for $%
i^{\prime}\neq i,f$ and $\langle\Phi_{f}|\Psi_{f}\rangle=1+O(|\langle%
\Phi_{f}|\Psi_{i}\rangle|^{2})$, we have, 
\begin{equation}
\langle\Phi_{f}|\nabla_{\mathbf{R}_{k}}\Psi_{i}\rangle=\frac{%
\langle\Psi_{f}|\nabla_{\mathbf{R}_{k}}H_{e}|\Psi_{i}\rangle}{\epsilon_{if}}%
\langle\Phi_{f}|\Psi_{f}\rangle=\frac{\langle\Psi_{f}|\nabla_{\mathbf{R}%
_{k}}H_{e}|\Psi_{i}\rangle}{\epsilon_{if}}.
\end{equation}
Similarly, 
\begin{equation}
\langle\nabla_{\mathbf{R}_{k}}\Phi_{f}|\Psi_{i}\rangle=-\frac{%
\langle\Phi_{f}|\nabla_{\mathbf{R}_{k}}H_{e}|\Phi_{i}\rangle}{\epsilon_{if}}%
\langle\Phi_{i}|\Psi_{i}\rangle=-\frac{\langle\Phi_{f}|\nabla_{\mathbf{R}%
_{k}}H_{e}|\Phi_{i}\rangle}{\epsilon_{if}}.  \label{leading}
\end{equation}
Combining these results and noting that $H_{1}^{BO}$ does not have diagonal
components, we arrive at 
\begin{equation}
\nabla_{\mathbf{R}_{k}}M_{e}^{BO}+\langle\Psi_{f}|\nabla_{\mathbf{R}%
_{k}}H|\Psi_{i}\rangle=\langle\Phi_{f}|\nabla_{\mathbf{R}_{k}}H|\Phi_{i}%
\rangle-\langle\Phi_{f}|\Psi_{i}\rangle\langle\Psi_{f}|\nabla_{\mathbf{R}%
_{k}}H|\Psi_{f}\rangle.
\end{equation}
We can use Eq. (\ref{FP}) and approximate $|\Psi_{f}\rangle\approx|\Phi_{f}%
\rangle$ to get Eq. (\ref{gradMe}).

\section{Evaluation of the overlap integral within the PAW}

\label{A3}

Consider the problem of evaluating the overlap integral $\langle\Psi|\Phi%
\rangle$ between two wave functions from two different solids (e.g., one is
a perfect crystal and the other contains a defect). Using the PAW expansion
of the full wave functions: 
\begin{equation}
|\Psi\rangle=|\tilde{\Psi}\rangle+|\Psi^{AE}\rangle_{a}-|\Psi^{PS}%
\rangle_{a},
\end{equation}
where $|\tilde{\Psi}\rangle$ is the pseudo wave function and $%
|\Psi^{AE}\rangle_{a}$ and $|\Psi^{PS}\rangle_{a}$ are the atomic wave
functions inside the augmentation sphere of each atom $a$, and similarly, 
\begin{equation}
|\Phi\rangle=|\tilde{\Phi}\rangle+|\Phi^{AE}\rangle_{b}-|\Phi^{PS}%
\rangle_{b}.
\end{equation}
Now, $\langle\Psi|\Phi\rangle$ is given as: 
\begin{eqnarray}
\langle\Psi|\Phi\rangle & = & \left(\langle\tilde{\Psi}|+_{a}\langle%
\Psi^{AE}|-_{a}\langle\Psi^{PS}|\right)\left(|\tilde{\Phi}%
\rangle+|\Phi^{AE}\rangle_{b}-|\Phi^{PS}\rangle_{b}\right)  \notag \\
& = & \langle\tilde{\Psi}|\tilde{\Phi}\rangle+\langle\tilde{\Psi}%
|\Phi^{AE}\rangle_{b}-\langle\tilde{\Psi}|\Phi^{PS}\rangle_{b}+_{a}\langle%
\Psi^{AE}|\tilde{\Phi}\rangle-_{a}\langle\Psi^{PS}|\tilde{\Phi}\rangle 
\notag \\
& + &
\left(_{a}\langle\Psi^{AE}|-_{a}\langle\Psi^{PS}|\right)\left(|\Phi^{AE}%
\rangle_{b}-|\Phi^{PS}\rangle_{b}\right).  \label{overlap}
\end{eqnarray}
The first term, $\langle\tilde{\Psi}|\tilde{\Phi}\rangle$, is the overlap of
the pseudo wavefunctions and can be easily calculated since the pseudo
wavefuntions are expanded in the same base set of plane waves.

In order to evaluate the terms $\langle\tilde{\Psi}|\Phi^{AE}\rangle_{b}-%
\langle\tilde{\Psi}|\Phi^{PS}\rangle_{b}$ and $_{a}\langle\Psi^{AE}|\tilde{%
\Phi}\rangle-_{a}\langle\Psi^{PS}|\tilde{\Phi}\rangle$, we make use of the
unitary operators constructed by the projectors $|\tilde{p}\rangle$ and the
pseudo atomic wavefunctions $|\tilde{\phi}\rangle$: 
\begin{equation}
\sum_{b,i_{b}}|\tilde{p}_{i_{b}}^{b}\rangle\langle\tilde{\phi}_{i_{b}}^{b}|=1
\end{equation}
and 
\begin{equation}
\sum_{a,i_{a}}|\tilde{\phi}_{i_{a}}^{a}\rangle\langle\tilde{p}_{i_{a}}^{a}|=1
\end{equation}
inside the augmentation sphere of each atom $b$ of the perfect crystal and
each atom $a$ of the solid with the defect respectively. Thus: 
\begin{eqnarray}
\langle\tilde{\Psi}|\Phi^{AE}\rangle_{b}-\langle\tilde{\Psi}%
|\Phi^{PS}\rangle_{b} & = & \sum_{b,i_{b}}\left(\langle\tilde{\Psi}|\tilde{p}%
_{i_{b}}^{b}\rangle\langle\tilde{\phi}_{i_{b}}^{b}|\Phi^{AE}\rangle_{b}-%
\langle\tilde{\Psi}|\tilde{p}_{i_{b}}^{b}\rangle\langle\tilde{\phi}%
_{i_{b}}^{b}|\Phi^{PS}\rangle_{b}\right)  \label{pcoverlap}
\end{eqnarray}
and 
\begin{eqnarray}
_{a}\langle\Psi^{AE}|\tilde{\Phi}\rangle-_{a}\langle\Psi^{PS}|\tilde{\Phi}%
\rangle & = & \sum_{a,{i_{a}}}\left(_{a}\langle\Psi^{AE}|\tilde{\phi}%
_{i_{a}}^{a}\rangle\langle\tilde{p}_{i_{a}}^{a}|\tilde{\Phi}%
\rangle-_{a}\langle\Psi^{PS}|\tilde{\phi}_{i_{a}}^{a}\rangle\langle\tilde{p}%
_{i_{a}}^{a}|\tilde{\Phi}\rangle\right)  \label{sdoverlap}
\end{eqnarray}

Equations (\ref{pcoverlap}) and (\ref{sdoverlap}) ensure that in the case
that if the two solids are identical, i.e. $|\tilde{\Psi}\rangle$ and $|%
\tilde{\Phi}\rangle$ are eigenstates of the same Hamiltonian and the
augmentations spheres are identical, the one center expansion $\sum_{i}|%
\tilde{\phi}\rangle\langle\tilde{p}|\tilde{\Psi}\rangle$ of the pseudo
wavefunction is identical to the pseudo wavefunction $|\tilde{\Psi}\rangle$
inside the augmentations sphere and 
\begin{equation}
\langle\tilde{\Psi}_{f}|\Phi^{AE}\rangle-\langle\tilde{\Psi}%
|\Phi^{PS}\rangle=\langle\tilde{\Psi}^{PS}|\Phi^{AE}\rangle-\langle\tilde{%
\Psi}^{PS}|\Phi^{PS}\rangle.
\end{equation}

To evaluate Eqs. (\ref{pcoverlap}) and (\ref{sdoverlap}), we need the
projections of the pseudo wavefunctions of the first solid to the projectors
of the atomic wavefunctions of the second solid, $\langle\tilde{\Psi}|\tilde{%
p}_{i_{b}}^{b}\rangle$, and vise versa for the projections $\langle\tilde{p}%
_{i_{a}}^{a}|\tilde{\Phi}\rangle$. This can be easily calculated since both
the pseudo wavefunctions and the projectors are expanded in the same base
set of plane waves.

The difficulty in evaluating the last term in Eq. (\ref{overlap}) $%
\left(_{a}\langle\Psi^{AE}|-_{a}\langle\Psi^{PS}|\right)\left(|\Phi^{AE}%
\rangle_{b}-|\Phi^{PS}\rangle_{b}\right)$ is that the cutoff spheres for the
two wave functions are usually not identical. We can bypass this difficulty
by evaluating the integral with the assistance of a complete set of plane
waves $|\mathbf{k}\rangle$, 
\begin{eqnarray}
\left(_{a}\langle\Psi^{AE}|-_{a}\langle\Psi^{PS}|\right)\left(|\Phi^{AE}%
\rangle_{b}-|\Phi_{i}^{PS}\rangle_{b}\right) & = & \sum_{\mathbf{k}%
}\left(_{a}\langle\Psi^{AE}|-_{a}\langle\Psi^{PS}|\right)|\mathbf{k}%
\rangle\langle\mathbf{k}|\left(|\Phi^{AE}\rangle_{b}-|\Phi^{PS}\rangle_{b}%
\right)  \notag \\
& = & \sum_{\mathbf{k}}\left(_{a}\langle\Psi^{AE}|\mathbf{k}%
\rangle-_{a}\langle\Psi^{PS}|\mathbf{k}\rangle\right)\left(\langle\mathbf{k}%
|\Phi^{AE}\rangle_{b}-\langle\mathbf{k}|\Phi^{PS}\rangle_{b}\right).  \notag
\end{eqnarray}
The plane waves can be expanded in either sphere as 
\begin{equation}
e^{i\mathbf{k}\cdot\mathbf{r}}=4\pi\sum_{lm}i^{l}j_{l}(kr)Y_{lm}^{\ast}(\hat{%
\mathbf{k}})Y_{lm}(\hat{\mathbf{r}}).
\end{equation}
and using, 
\begin{equation}
|\mathbf{k}\rangle=\frac{1}{\sqrt{V}}e^{i\mathbf{k}\cdot\mathbf{r}},
\end{equation}
the all-electron and the pseudo atomic wave functions is written as: 
\begin{equation}
|\Phi^{AE}\rangle_{b}=\sum_{b,i_{b}}R_{b,i_{b}}^{AE}Y_{l_{b},m_{b}}\langle%
\tilde{p}_{b,i_{b}}|\tilde{\Phi}_{i}\rangle,
\end{equation}
\begin{equation}
|\Phi^{PS}\rangle_{b}=\sum_{b,i_{b}}R_{b,i_{b}}^{PS}Y_{l_{b},m_{b}}\langle%
\tilde{p}_{b,i_{b}}|\tilde{\Phi}_{i}\rangle,
\end{equation}
\begin{equation}
_{a}\langle\Psi^{AE}|\mathbf{k}\rangle-_{a}\langle\Psi^{PS}|\mathbf{k}%
\rangle=\frac{4\pi}{\sqrt{V}}\sum_{a,i_{a}}\langle\tilde{\Psi}|\tilde{p}%
_{a,i_{a}}\rangle e^{i\mathbf{k}\cdot\mathbf{R_{a}}%
}i^{l_{a}}Y_{l_{a},m_{a}}^{\ast}(\hat{\mathbf{k}}%
)\int_{0}^{r_{a}}j_{l_{a}}(kr)(R_{a,i_{a}}^{AE}-R_{a,i_{a}}^{PS})r^{2}dr,
\end{equation}
and 
\begin{equation}
\langle\mathbf{k}|\Phi^{AE}\rangle_{b}-\langle\mathbf{k}|\Phi^{PS}%
\rangle_{b}=\frac{4\pi}{\sqrt{V}}\sum_{b,i_{b}}\langle\tilde{p}_{b,i_{b}}|%
\tilde{\Phi}\rangle e^{-i\mathbf{k}\cdot\mathbf{R_{b}}%
}(-i)^{l_{b}}Y_{l_{b},m_{b}}(\hat{\mathbf{k}}%
)\int_{0}^{r_{b}}j_{l_{b}}(kr)(R_{b,i_{b}}^{AE}-R_{b,i_{b}}^{PS})r^{2}dr,
\end{equation}

\section{Phonon integrals}

\label{A4}

The overlap matrix between the initial and final states for the mode $j$ is, 
\begin{equation}
\langle X_{n_{j}^{f}}(q_{j}+\delta q_{j})|X_{n_{j}^{i}}(q_{j})\rangle=\int
X_{n_{j}+p_{j}}(q_{j}+\delta q_{j})X_{n_{j}}(q_{j})dq_{j}.
\label{eq:overlap}
\end{equation}
where $n_{j}^{i}=n_{j}$ and $n_{j}^{f}=n_{j}+p_{j}$.

For convenience, we drop the subscript $j$ for $n_{j}$ and $p_{j}$.
Expanding $X_{n}(q_{j}+\delta q_{j})$ in terms of $\delta q_{j}$, 
\begin{equation}
X_{n}(q_{j}+\delta q_{j})=\sum_{l}\frac{1}{l!}\frac{d^{l}X_{n}(q_{j})}{%
dq_{j}^{l}}\delta q_{j}^{l}.
\end{equation}
Defining the raising and lowering operators 
\begin{equation}
\hat{a}_{\pm}=\mp\sqrt{\frac{\hbar}{2\omega_{j}}}\frac{d}{dq_{j}}+\sqrt{%
\frac{\omega_{j}}{2\hbar}}q_{j},
\end{equation}
we have, 
\begin{equation}
\hat{a}_{+}X_{n}(q_{j})=\sqrt{n+1}X_{n+1}(q_{j}),
\end{equation}
and 
\begin{equation}
\hat{a}_{-}X_{n}(q_{j})=\sqrt{n}X_{n-1}(q_{j}).
\end{equation}
Subtracting the two, we find, 
\begin{equation}
\frac{d}{dq_{j}}X_{n}(q_{j})=\sqrt{\frac{\omega_{j}}{2\hbar}}(\hat{a}_{-}-%
\hat{a}_{+})X_{n}(q_{j})=\sqrt{\frac{n\omega_{j}}{2\hbar}}X_{n-1}(q_{j})-%
\sqrt{\frac{(n+1)\omega_{j}}{2\hbar}}X_{n+1}(q_{j}).
\end{equation}
Using this recursive relation, we find that the lowest order term for $\int
X_{n}(q_{j}+\delta q_{j})X_{n+k}(q_{j})dq_{j}$ is $\delta q_{j}^{|k|}$.
Therefore, for small $\delta q{}_{j}$ only $k=\pm1$ terms dominates. It
means that each mode would at most emit or absorb a single phonon.

The result for the integrals are, 
\begin{equation}
\int\frac{dX_{n}(q)}{dq}X_{n+1}(q)dq=-\sqrt{\frac{(n+1)\omega_{j}}{2\hbar}},
\end{equation}
(note that this was incorrectly given as $-(\sqrt{\hbar m\omega/2})\sqrt{n+1}
$ in Ref. \onlinecite{Huang:1950}), and, 
\begin{equation}
\int\frac{dX_{n}(q)}{dq}X_{n-1}(q)dq=\sqrt{\frac{n\omega_{j}}{2\hbar}}.
\end{equation}

For linear phonon matrix elements, we have, 
\begin{equation}
q_{j}X_{n}(q_{j})=\sqrt{\frac{\hbar}{2\omega_{j}}}(\hat{a}_{-}+\hat{a}%
_{+})X_{n}(q_{j})=\sqrt{\frac{n\hbar}{2\omega_{j}}}X_{n-1}(q_{j})+\sqrt{%
\frac{(n+1)\hbar}{2\omega_{j}}}X_{n+1}(q_{j}).
\end{equation}
The integrals needed are, 
\begin{equation}
\int X_{n}(q)X_{n+1}(q)qdq=\sqrt{\frac{(n+1)\hbar}{2\omega_{j}}},
\end{equation}

\begin{equation}
\int X_{n}(q)X_{n-1}(q)qdq=\sqrt{\frac{n\hbar}{2\omega_{j}}},
\end{equation}
and, 
\begin{equation}
\int\frac{dX_{n}(q)}{dq}X_{n}(q)qdq=-\frac{1}{2}.
\end{equation}

\section{Line shape function}

\label{A5}

We first consider a single phonon band, i.e., all of the phonon modes $%
\omega_{j}=\omega(\mathbf{k}_{j})$ form a single continuous band described
by wave vectors $\mathbf{k}_{j}$. Because of the Born-von-Karman periodic
boundary condition, the phonon band is discretized into $N$ modes. Suppose
that $s$ modes go down by one quantum and $s+p$ modes go up by one quantum.
Then the line shape factor, Eq. (\ref{eq:LSFproduct}) with $M=1$, contains
contributions formed from the following products, 
\begin{equation}
\left\{ \prod_{j=1}^{N}t_{j}\right\} \left\{ \prod_{k\in s}f_{k,-}\right\}
\left\{ \prod_{l\in s+p}f_{l+}\right\} \times  \notag
\end{equation}

\begin{equation}
\times\delta\left(\sum_{l\in
s+p}[n_{l}^{i}\hbar(\omega_{l}^{f}-\omega_{l}^{i})+\hbar\omega_{l}^{f}]+%
\sum_{k\in
s}[n_{k}^{i}\hbar(\omega_{k}^{f}-\omega_{k}^{i})-\hbar\omega_{k}^{f}]+%
\sum_{m\ni\{s,s+p\}}n_{m}^{i}\hbar(\omega_{m}^{f}-\omega_{m}^{i})+%
\epsilon_{if}\right),  \notag
\end{equation}

\begin{equation}
\left\{ \prod_{j=1}^{N}t_{j}\right\} \left\{ \prod_{k\in s}f_{k,-}\right\}
\left\{ \prod_{l\in s+p}f_{l+}\right\} \delta\left(\sum_{l\in
s+p}\hbar\omega_{l}^{f}-\sum_{k\in
s}\hbar\omega_{k}^{f}+\sum_{j=1}^{N}n_{j}^{i}\hbar(\omega_{j}^{f}-%
\omega_{j}^{i})+\epsilon_{if}\right),
\end{equation}
where $\sum_{j=1}^{N}n_{j}^{i}\hbar(\omega_{j}^{f}-\omega_{j}^{i})$ is the
energy difference because of the different phonon frequencies of the initial
and final configuration of the defect and $t_{j}$, $f_{-}$, and $f_{+}$ are
defined as: 
\begin{align}
t_{j} & =\left|\int X_{n_{j}}(q_{j})X_{n_{j}}(q_{j}+\delta
q_{j})dq_{j}\right|^{2}  \notag \\
f_{k,-} & =\frac{{\displaystyle \left|\int
X_{n_{k}}(q_{k})X_{n_{k}-1}(q_{k}+\delta q_{k})dq_{k}\right|^{2}}}{{%
\displaystyle \left|\int X_{n_{k}}(q_{k})X_{n_{k}}(q_{k}+\delta
q_{k})dq_{k}\right|^{2}}} \\
f_{l,+} & =\frac{{\displaystyle \left|\int
X_{n_{l}}(q_{l})X_{n_{l}+1}(q_{l}+\delta q_{l})dq_{l}\right|^{2}}}{{%
\displaystyle \left|\int X_{n_{l}}(q_{l})X_{n_{l}}(q_{l}+\delta
q_{l})dq_{l}\right|^{2}}}.  \notag
\end{align}
A naive way to sum over all possible configurations is to neglect the
difference in the frequencies and apply the same counting method as Huang
and Rhys \cite{Huang:1950} to write the configurational sum for all such
combinations of phonons as, 
\begin{equation}
\frac{1}{s!(s+p)!}\left\{ \prod_{j=1}^{N}t_{j}\right\} \left\{
\sum_{k=1}^{N}f_{k,-}\right\} ^{s}\left\{ \sum_{l=1}^{N}f_{k,+}\right\}
^{s+p}\delta\left(\sum_{l\in s+p}\hbar\omega_{l}^{f}-\sum_{k\in
s}\hbar\omega_{k}^{f}+\sum_{j=1}^{N}n_{j}^{i}\hbar(\omega_{j}^{f}-%
\omega_{j}^{i})+\epsilon_{if}\right).
\end{equation}
This would not be correct if the frequencies are different for each mode.
Furthermore, the summation over configurations for large $N$ is needed to
integrate out the $\delta$ function. Therefore the $\delta$ function cannot
be left outside the summations. Let us consider one term in the $\delta$
function at a time. Consider one the plus terms $\hbar\omega_{m}^{f}$ and
insert the $\delta$ function into one of the summations, 
\begin{eqnarray}
& & \frac{1}{s!(s+p)!}\left\{ \prod_{j=1}^{N}t\right\} \left\{
\sum_{k=1}^{N}f_{k,-}\right\} ^{s}\left\{ \sum_{l=1}^{N}f_{k,+}\right\}
^{s+p-1}  \notag \\
& & \sum_{m=1}^{N}\left\{ f_{m,+}\delta\left(\hbar\omega_{m}^{f}+\sum_{l\in
s+p-1}\hbar\omega_{l}^{f}-\sum_{k\in
s}\hbar\omega_{k}^{f}+\sum_{j=1}^{N}n_{j}^{i}\hbar(\omega_{j}^{f}-%
\omega_{j}^{i})+\epsilon_{if}\right)\right\} .  \label{eq:HuangLSF}
\end{eqnarray}
For large $N$, each of the summations inside the curly brackets can be
converted into integrals and evaluated, 
\begin{equation}
S_{\pm}=\sum_{k=1}^{N}f_{k,\pm}=\frac{N}{\Omega_{\mathbf{k}}}\int f_{\mathbf{%
k},\pm}d\mathbf{k}=%
\begin{pmatrix}
n+1 \\ 
n%
\end{pmatrix}%
\frac{\omega}{2\hbar}N\delta q^{2},  \label{eq:Spm}
\end{equation}
where $\Omega_{\mathbf{k}}$ is the volume of the reciprocal space Brillouin
zone. In the last step we assumed that the frequency and displacement do not
change with $\mathbf{k}$.

In order to evaluated the last factor that includes the $\delta$ function,
we note that each term in the summation over $m$ has a different $%
\omega_{m}^{f}$, which spans the entire phonon band when $m$ scans from $1$
to $N$. Thus as we convert the sum over $m$ to integral over $\mathbf{k}$,
the argument $\omega_{m}^{f}$ is also converted to $\omega_{\mathbf{k}}$, 
\begin{eqnarray}
& & \sum_{m=1}^{N}\left\{ f_{m,+}\delta\left(\hbar\omega_{m}^{f}+\sum_{l\in
s+p-1}\hbar\omega_{l}^{f}-\sum_{k\in
s}\hbar\omega_{k}^{f}+\sum_{j=1}^{N}n_{j}^{i}\hbar(\omega_{j}^{f}-%
\omega_{j}^{i})+\epsilon_{if}\right)\right\}  \notag \\
& & \approx\frac{N}{\Omega_{\mathbf{k}}}\int f_{\mathbf{k}%
,+}\delta\left(\hbar\omega_{\mathbf{k}}^{f}+\sum_{l\in
s+p-1}\hbar\omega_{l}^{f}-\sum_{k\in
s}\hbar\omega_{k}^{f}+\sum_{j=1}^{N}n_{j}^{i}\hbar(\omega_{j}^{f}-%
\omega_{j}^{i})+\epsilon_{if}\right)d\mathbf{k}  \notag \\
& & =\left.S_{+}\frac{D(\omega)}{\Omega_{\mathbf{k}}}\right|_{\hbar%
\omega^{f}+\sum_{l\in s+p-1}\hbar\omega_{l}^{f}-\sum_{k\in
s}\hbar\omega_{k}^{f}+\sum_{j=1}^{N}n_{j}^{i}\hbar(\omega_{j}^{f}-%
\omega_{j}^{i})+\epsilon_{if}=0},  \label{eq:phononDOS}
\end{eqnarray}
where $D(\omega)$ is the phonon density of states. Combining the above
equations and then setting all frequencies to $\omega$, Eq. (\ref%
{eq:HuangLSF}) now becomes, 
\begin{equation}
\frac{1}{s!(s+p)!}\left\{ \prod_{j=1}^{N}t\right\} S_{-}^{s}S_{+}^{s+p}\left.%
\frac{D(\omega)}{\Omega_{\mathbf{k}}}\right|_{p\hbar\omega+%
\sum_{j=1}^{N}n_{j}^{i}\hbar(\omega_{j}^{f}-\omega_{j}^{i})+\epsilon_{if}=0}.
\end{equation}
But there is one such contribution for each $\omega_{k}$ or $\omega_{l}$ in
the $\delta$ function, regardless of the sign of the frequency. For $s$
modes subtracting a phonon and $s+p$ modes adding a phonon there are total $%
2s+p$ such contributions. We thus sum over all the terms and obtain, 
\begin{equation}
\frac{2s+p}{s!(s+p)!}\left\{ \prod_{j=1}^{N}t_{j}\right\}
S_{-}^{s}S_{+}^{s+p}\left.\frac{D(\omega)}{\Omega_{\mathbf{k}}}%
\right|_{p\hbar\omega+\sum_{j=1}^{N}n_{j}^{i}\hbar(\omega_{j}^{f}-%
\omega_{j}^{i})+\epsilon_{if}=0}.
\end{equation}
Finally, the factor $\prod_{j=1}^{N}t_{j}$ is, 
\begin{equation}
\prod_{j=1}^{N}\left|\int X_{n_{j}}(q_{j})X_{n_{j}}(q_{j}+\delta
q_{j})dq_{j}\right|^{2}=\left[1-\frac{(2n+1)\omega}{4\hbar}\delta q^{2}%
\right]^{2N}=\exp\left[-(S_{+}+S_{-})\right].
\end{equation}
The line shape factor for a single phonon band is, 
\begin{eqnarray}
& & \left.\frac{D(\omega)}{\Omega_{\mathbf{k}}}\right|_{p\hbar\omega+%
\epsilon_{if}=0}\exp\left[-(S_{+}+S_{-})\right]\sum_{s=0}^{\infty}\frac{2s+p%
}{s!(s+p)!}S_{+}^{s+p}S_{-}^{s}  \notag \\
& = & \left.\frac{D(\omega)}{\Omega_{\mathbf{k}}}\right|_{p\hbar\omega+%
\epsilon_{if}=0}\exp\left[-(S_{+}+S_{-})\right]\left(\frac{S_{+}}{S_{-}}%
\right)^{p/2}\times  \notag \\
& & \left[pI_{p}\left(2\sqrt{S_{+}S_{-}}\right)+2\sqrt{S_{+}S_{-}}%
I_{p+1}\left(2\sqrt{S_{+}S_{-}}\right)\right].
\end{eqnarray}

To generalize the above expression to multiple phonon bands, the
normalization factor must be evaluated with a summation over both the band
index and the $\mathbf{k}$ points within each band. If we use $F_{j}$ to
denote the factor for a band that adds net $p_{j}$ phonons, i.e., 
\begin{equation}
F_{j}=\sum_{s_{j}=0}^{\infty}\frac{1}{s_{j}!(s_{j}+p_{j})!}\left\{
\prod_{m=1}^{N}t_{jm}\right\} \left\{ \sum_{k=1}^{N}f_{jk,-}\right\}
^{s_{j}}\left\{ \sum_{l=1}^{N}f_{jl,+}\right\} ^{s_{j}+p_{j}}
\end{equation}
then in a similar manner as for the case of a single phonon band, $F_{j}$ is
evaluated to be, 
\begin{equation}
F_{j}=\left(\frac{n_{j}+1}{n_{j}}\right)^{p_{j}/2}\exp\left[-S_{j}(2n_{j}+1)%
\right]I_{p_{j}}\left[2S_{j}\sqrt{n_{j}(n_{j}+1)}\right].
\end{equation}

Now we insert the $\delta$ function into the product of $F_{j}$ in the same
manner as in the case of a single band to form the full line shape factor,
one phonon band at a time. For now let us consider the case where all $p_{j}$%
's are positive. We have, 
\begin{eqnarray}
& & \frac{\prod_{j=1}^{M}F_{j}}{F_{j^{\prime\prime}}}\sum_{s_{j^{\prime%
\prime}}=0}^{\infty}\frac{2s_{j^{\prime\prime}}+p_{j^{\prime\prime}}}{%
s_{j^{\prime\prime}}!(s_{j^{\prime\prime}}+p_{j^{\prime\prime}})!}\left\{
\prod_{m=1}^{N}t_{j^{\prime\prime}m}\right\} \left\{
\sum_{k=1}^{N}f_{j^{\prime\prime}k,-}\right\} ^{s_{j^{\prime\prime}}}\left\{
\sum_{l=1}^{N}f_{j^{\prime\prime}l,+}\right\}
^{s_{j^{\prime\prime}}+p_{j^{\prime\prime}}-1}\times  \notag \\
& &
\sum_{m=1}^{N}f_{j^{\prime\prime}m,+}\delta\left(\hbar\omega_{j^{\prime%
\prime}m}^{f}+\sum_{l\in
s_{j^{\prime\prime}}+p_{j^{\prime\prime}}-1}\hbar\omega_{j^{\prime%
\prime}l}^{f}-\sum_{k\in
s_{j^{\prime\prime}}}\hbar\omega_{j^{\prime\prime}k}^{f}+\sum_{l^{%
\prime}=1}^{N}n_{j^{\prime\prime}l^{\prime}}^{i}\hbar(\omega_{j^{\prime%
\prime}l^{\prime}}^{f}-\omega_{j^{\prime\prime}l^{\prime}}^{i})+\right. 
\notag \\
& & \left.\sum_{j^{\prime}\neq j^{\prime\prime},l\in
s_{j^{\prime}}+p_{j^{\prime}}}\hbar\omega_{j^{\prime}l}^{f}-\sum_{j^{\prime}%
\neq j^{\prime\prime},l\in
s_{j^{\prime}}}\hbar\omega_{j^{\prime}l}^{f}+\sum_{j^{\prime}\neq
j^{\prime\prime},k^{\prime}=1}^{N}n_{j^{\prime}k^{\prime}}^{i}\hbar(%
\omega_{j^{\prime}k^{\prime}}^{f}-\omega_{j^{\prime}k^{\prime}}^{i})+%
\epsilon_{if}\right)  \notag \\
& = & \left(\prod_{j=1}^{M}F_{j}\right)\left.\frac{D(\omega_{j^{\prime%
\prime}})}{\Omega_{\mathbf{k}}}\right|_{\sum_{j^{\prime}}p_{j^{\prime}}\hbar%
\omega_{j^{\prime}}+\sum_{j=1}^{Md}\sum_{l=1}^{N}n_{jl}^{i}\hbar(%
\omega_{jl}^{f}-\omega_{jl}^{i})+\epsilon_{if}=0}\times  \notag \\
& & \left\{ p_{j^{\prime\prime}}+2S_{j^{\prime\prime}}\sqrt{%
n_{j^{\prime\prime}}(n_{j^{\prime\prime}}+1)}\frac{I_{p_{j^{\prime\prime}}+1}%
\left[2S_{j^{\prime\prime}}\sqrt{n_{j^{\prime\prime}}(n_{j^{\prime\prime}}+1)%
}\right]}{I_{p_{j^{\prime\prime}}}\left[2S_{j^{\prime\prime}}\sqrt{%
n_{j^{\prime\prime}}(n_{j^{\prime\prime}}+1)}\right]}\right\} .
\label{positivep}
\end{eqnarray}
where $j^{\prime\prime}$ is one of the phonon bands and we have used Eqs. (%
\ref{eq:Spm}) and (\ref{eq:phononDOS}). Summing over all possible $%
j^{\prime\prime}$ terms and with an additional summation over all
configurations $\{p_{j}\}$, we find, 
\begin{equation}
F=\frac{1}{\Omega_{\mathbf{k}}}\sum_{\{p_{j}\}}\left.\left\{
\left(\prod_{j=1}^{M}F_{j}\right)\sum_{j=1}^{M}\left\{ p_{j}+2S_{j}\sqrt{%
n_{j}(n_{j}+1)}\frac{I_{p_{j}+1}\left[2S_{j}\sqrt{n_{j}(n_{j}+1)}\right]}{%
I_{p_{j}}\left[2S_{j}\sqrt{n_{j}(n_{j}+1)}\right]}\right\}
D(\omega_{j})\right\}
\right|_{\sum_{j=1}^{M}p_{j}\hbar\omega_{j}+\epsilon_{if}=0}.  \label{F1}
\end{equation}
If some of the $p_{j}$'s are negative, we need to switch the roles of $S_{+}$
and $S_{-}$ following Ref. \onlinecite{Huang:1950}. Redefining $%
s_{j}+p_{j}\to s_{j}$ and $s_{j}\to s_{j}-p_{j}$ in Eq. (\ref{positivep}),
the factor corresponding to $p_{j}$ becomes, 
\begin{eqnarray}
& & -p_{j}+2S_{j}\sqrt{n_{j}(n_{j}+1)}\frac{I_{-p_{j}+1}\left[2S_{j}\sqrt{%
n_{j}(n_{j}+1)}\right]}{I_{-p_{j}}\left[2S_{j}\sqrt{n_{j}(n_{j}+1)}\right]} 
\notag \\
& = & p_{j}+2S_{j}\sqrt{n_{j}(n_{j}+1)}\frac{I_{p_{j}+1}\left[2S_{j}\sqrt{%
n_{j}(n_{j}+1)}\right]}{I_{p_{j}}\left[2S_{j}\sqrt{n_{j}(n_{j}+1)}\right]},
\end{eqnarray}
using the recurrence relation for the Bessel functions. Therefore Eq. (\ref%
{F1}) is valid for both positive and negative $p_{j}$'s. Applying
thermodynamic average to the occupation numbers, $n_{j}$ is replaced by the
Bose-Einstein distribution function, 
\begin{equation}
n_{j}\to\frac{1}{\exp(\hbar\omega_{j}/kT)-1},
\end{equation}

\begin{equation}
\frac{n_{j}+1}{n_{j}}\to\exp\left(\frac{\hbar\omega}{kT}\right),
\end{equation}

\begin{equation}
2n_{j}+1\to\coth\left(\frac{\hbar\omega}{2kT}\right),
\end{equation}
and 
\begin{equation}
2\sqrt{n_{j}(n_{j}+1)}\to\frac{1}{\sinh(\hbar\omega/2kT)},
\end{equation}
we obtain Eqs. (\ref{Fs}) and (\ref{F}).

\end{document}